\documentclass{nature}
\usepackage{graphicx,epsfig,amssymb,amsmath}
\usepackage{epstopdf}
\usepackage{lineno}
\usepackage{caption}
\def\gsim{\mathrel{\raise.5ex\hbox{$>$}\mkern-14mu
             \lower0.6ex\hbox{$\sim$}}}
\def\lsim{\mathrel{\raise.3ex\hbox{$<$}\mkern-14mu
             \lower0.6ex\hbox{$\sim$}}}

\def\gsim{\mathrel{\raise.5ex\hbox{$>$}\mkern-14mu
             \lower0.6ex\hbox{$\sim$}}}
\def\lsim{\mathrel{\raise.3ex\hbox{$<$}\mkern-14mu
             \lower0.6ex\hbox{$\sim$}}}
\def\gro{GRO~J1655-40}
\def\Msun{M_\odot}

\def\mathv{\textbf{\em V}}
\def\mathB{\textbf{\em B}}

\def\mathJ{\textbf{\em J}}
\def\gsim{\mathrel{\raise.5ex\hbox{$>$}\mkern-14mu
             \lower0.6ex\hbox{$\sim$}}}
\def\lsim{\mathrel{\raise.3ex\hbox{$<$}\mkern-14mu
             \lower0.6ex\hbox{$\sim$}}}
\def\gro{GRO~J1655-40}
\def\deg{^\circ}
\def\aa{\buildrel _{\circ} \over {\mathrm{A}}}
\def\Msun{M_\odot}

\def\fexxiv{Fe\,{\sc xxiv}}
\def\fexxv{Fe\,{\sc xxv}}
\def\fexxvi{Fe\,{\sc xxvi}}
\def\fexxiii{Fe\,{\sc xxiii}}
\def\fexxii{Fe\,{\sc xxii}}

\def\mnxxv{Mn\,{\sc xxv}}
\def\mnxxiv{Mn\,{\sc xxiv}}
\def\crxxiv{Cr\,{\sc xxiv}}
\def\crxxiii{Cr\,{\sc xxiii}}

\def\nixxviii{Ni\,{\sc xxviii}}

\def\arxviii{Ar\,{\sc xviii}}
\def\arxvii{Ar\,{\sc xvii}}

\def\sxvi{S\,{\sc xvi}}
\def\sxv{S\,{\sc xv}}
\def\sixiv{Si\,{\sc xiv}}
\def\caxx{Ca\,{\sc xx}}
\def\caxix{Ca\,{\sc xix}}
\def\mgxii{Mg\,{\sc xii}}
\def\nex{Ne\,{\sc x}}

\def\mgxii{Mg\,{\sc xii}}

\title{%THE UNIVERSAL MAGNETIC STRUCTURE OF BLACK HOLE ACCRETION DISK WINDS
Magnetic Origin of Black Hole Winds Across the Mass Scale}

\author{Keigo Fukumura$^{1}$, Demosthenes Kazanas$^2$, Chris Shrader$^{2,3}$,
Ehud Behar$^{4,5}$, Francesco Tombesi$^{2,5,6}$ \& Ioannis Contopoulos$^7$}

\begin{document}

\maketitle

\begin{affiliations}
 \item James Madison University, Harrisonburg, VA 22807
 \item NASA, Goddard Space Flight Center, Greenbelt, MD 20771
 \item Universities Space Research Association, 10211 Wincopin Cir, Suite 500, Columbia, MD 21044, USA
 \item Physics Department, Technion, Haifa 32000, Israel
 \item Astronomy Department, University of Maryland, College Park, MD
 20742
 \item Dipartimento di Fisica, Universita'di Roma Tor Vergata, Via della Ricerca Scientifica 1, I-00133 Roma, Italy
 \item Academy of Athens, Soranou Efesiou 2, GR 11527 Athens, Greece
% \item separate with \verb|\item| commands.
\end{affiliations}

\begin{abstract}
\textbf{Black hole accretion disks appear to produce invariably plasma outflows that
result in blue-shifted absorption features in their spectra$^1$. The X-ray
absorption-line properties of these outflows are quite diverse, ranging in velocity
from non-relativistic$^2$ ($\sim 300$ km/sec) to sub-relativistic$^3$ ($\sim 0.1c$
where $c$ is the speed of light) and a similarly broad range in the ionization states
of the wind plasma$^{2,4}$.
We report here that semi-analytic, self-similar magnetohydrodynamic (MHD) wind models
that have successfully accounted for the X-ray absorber properties of supermassive black
holes$^{5,6}$, also fit well the high-resolution X-ray spectrum of the accreting
stellar-mass black hole, GRO J1655-40.
This provides an explicit theoretical argument of their MHD origin (aligned with
earlier observational claims)$^{7}$ and supports the notion of a universal magnetic
structure of the observed winds across all known black hole sizes}.
\end{abstract}

%Accretion disk winds are a ubiquitous feature of accreting black holes, manifesting in their blue-shifted spectral absorption lines.
%
The importance of these outflows (winds) lies partly in the fact that they may
remove the accretion disks' angular momentum$^{8}$, a condition necessary for
making accretion possible, and partly on their potential feedback and influence on
their environment$^{9}$. The variety of processes that can launch such
winds$^{10,11,12,13,14,15,43}$ has made their origin contentious. However, X-ray
spectroscopic data and analysis of the wind associated with the  X-ray binary (XRB)
GRO J1655-40 $^{7,16,17,18}$ have argued in favor of a magnetic origin, but only by
excluding the other candidate processes. Here, we present first-principles
computations of absorption line spectra of self-similar MHD accretion disk wind
models$^{5,6}$ that reproduce both the combined global ionization properties and
also the detailed kinematic structure of the absorption features of GRO J1655-40.
Most importantly, these wind models are the same as those that have accounted
successfully for the X-ray absorber systematics of active galactic nuclei (AGNs) as
diverse as Seyfert galaxies$^{5,19}$ and broad absorption line (BAL) quasars$^{6}$,
modified only by the different ionizing spectrum of GRO J1655-40 and scaled to its
black hole mass.
This demonstrates a universality of accretion disk wind properties across the
entire (10 $\mathbf{M_{\odot}-} {10^9} \mathbf{M_{\odot}})$ black hole mass
range$^{20}$.

The earlier observational claims favoring the magnetic origin of the \gro\ winds
were made by excluding the other two plausible mechanisms for launching winds off
accretion disks on the basis of the observed line properties of this
system$^{7,16}$; These are: (i) Thermally driven winds$^3$, launched by heating the
disk surface by the X-rays produced near the compact object to the Compton
temperature $T_C$. For a sufficiently large distance $R_C$ along the disk, the
plasma thermal velocity $V_{th}$ can be larger than the local escape velocity,
$kT_C/m_p \simeq V_{th}^2 > V_e^2/2 \sim GM/R_C$ ($G$ is the gravitational constant
and $M, m_p$ the black hole and proton masses respectively), allowing escape of the
X-ray heated matter to infinity. The GRO J1655-40 quasi-thermal disk spectrum of
$kT \sim 1.34$ keV, then, implies$^2$ $R_C > 7 \times 10^{12}$ cm, larger than the
binary orbit and grossly inconsistent with the photoionized wind properties
(however see$^{21,22,23}$). (ii) Radiation pressure driven winds (at sub-Eddington
luminosity, i.e. $L<L_{\rm Edd}$); these rely on the increased line pressure on
partially ionized elements, in analogy with stellar winds$^{24}$. However, as also
indicated earlier$^{7}$, the high values of the ionization parameter $\xi = L_X/(n
r^2)$ ($L_X$ is the ionizing luminosity, $n$ the plasma density and $r$ the radial
distance from the ionizing source) implied by the X-ray continuum and line
luminosity of GRO J1655-40 ($\xi > 10^3$) argue for an over-ionized wind plasma,
thereby excluding line pressure driven winds$^{25}$.

Winds can also be launched with the help of magnetic fields, either in combination
with the disk centrifugal forces$^{12,13,15}$ or by the disk vertical magnetic
pressure gradients$^{14}$. Their morphological distinction from those launched by
the other two processes lies in their inherently 2D structure, as they are launched
across the entire disk domain. Because of the large dynamic range of their
launching radii ($R_{\rm disk}/R_{S} \sim 10^6 ; R_S$ is the Schwarzschild radius),
they are best described by self-similar models with their density and velocity
spanning similarly large dynamic ranges. This fact is consistent with the wide
$\xi-$range of the ions observed in the X-ray absorption line spectra of AGN.
Self-similarity allows separation of the wind density $n(r,\theta)$ in $r$ (the
spherical radial coordinate) and $\theta$ (the disk polar angle), i.e.
$n(r,\theta)= n_0 f(\theta) (r/r_0)^{-(1+\alpha)}$ with$^5$ $n_0 \propto \dot m/M$
($M$ is the black hole mass and $\dot m \equiv \dot M/\dot M_{\rm Edd}, M_{\rm
Edd}= L_{\rm Edd}/c^2$) and $f(\theta) \simeq e^{5(\theta - \pi/2)}$,
giving these winds a toroidal appearance (see Fig.~1). { Note here that $n_0$ is
the wind density at the innermost launching radius at $r=r_0$.} With $r$ measured
in units of the Schwarzschild radius $R_S(\propto M)$, the wind column $N_H \propto
n(r) \,r \propto r^{-\alpha}$ is independent of the mass $M$, as also are  $\xi
\propto L/n(r) r^2 \propto r^{-(1-\alpha)}$ (assuming $L \propto \dot m M$) and
$V/c \propto r^{-1/2}$, with $n_0,\alpha$ the main free model parameters. The {\bf
computed} radial profiles of $N_H, V {\rm and}~ \xi$ for $\alpha = 0.2$ are shown
in Figure~2.

With the physical parameters that determine the winds' spectroscopic properties,
$N_H, V, \xi$, independent of the compact object mass, $M$, these models are
equally well applicable to AGNs and XRBs, the greatest differences between these
two classes being the spectral energy distribution (SED) of ionizing
radiation$^{6}$. Eliminating $r$ between the Hydrogen equivalent column of an ion,
$N_H$, and its $\xi-$value we obtain $N_H(\xi) \propto \xi^{\alpha/(1-\alpha)}$,
the Absorption Measure Distribution (AMD)$^{4,26}$; this relation can provide the
wind density radial profile, i.e. the value of  $\alpha$, by observations of
absorption line properties alone. The crucial test of any wind model is then to
produce the correct values of $N_H$ {\sl and} $V$ for the values of $\xi$ at which
the observed ions are present, {\sl for all ionic species}. Most works$^{16,27}$
modeling the X-ray absorbers of GRO J1655-40 focus on fitting accurately the
\fexxv-\fexxvi\ profiles to obtain the local values of $N_H,\xi$ and $V$. { In
contrast, our approach employs an established MHD wind model and aims, by
reproducing the  combined ionization/kinematic properties of the {\sl ensemble} of
observed transitions, to determine its global parameters $n_0$ and $\alpha$. }

%We have computed the photoionization structure of the self-similar MHD wind models
%outlined above following our own procedure$^{5,6,19}$ for different values of the
%parameter $\alpha$, with $n_0$ the value of the wind density at $\theta = \pi/2$ and
%$r_0 \simeq 2 R_S$, chosen so that $n_0 r_0 \sigma_T \simeq 3$ ($\sigma_T$ is the
%Thomson cross section). The input continuum spectrum was a multicolor disk with
%innermost temperature $kT = 1.34 \;{\rm keV}^{16}$. We have used several values
%of $\alpha$ with best fit for $\alpha = 0.2$. The results of these calculations are
%shown in figure 3 that provide the spectrum between 1.5 \AA \, and 12.2 \AA.

We have computed the photoionization structure of self-similar MHD winds following
our own procedure$^{5,6,19}$ with input continuum spectrum a multicolor disk of
innermost temperature $kT = 1.34 \;{\rm keV}^{16}$. { Because the procedure is
computationally intensive, $\alpha$ and $n_0$ were not varied independently but in
synchrony so that they produce an AMD consistent broadly with observations (see
further discussion in Methods). We have thus used a grid of representative values
for $\alpha$ (as listed in Table~S1) with which the broad-band fitting yields the
best-fit density normalization of $\tilde{n}_{\rm 17}=9.3$ (or $n_0 \simeq 10^{18}$
cm$^{-3}$) with $\alpha=0.2$ wind. } The results of these calculations are shown in
Figure 3 that provide the best-fit spectrum between 1.5 \AA \, and 12.2 \AA\ {\bf
based on the comparison between modeled and observed equivalent width (EW) for
major lines analyzed in this work (see Fig.~S4 in Methods for details).} %%

Our photoionization calculations split the radial coordinate $r$ along the LoS into
6-7 slabs per decade with $\Delta r/r \simeq 0.15$ and employ XSTAR$^{28}$ to
compute the local ionic abundances and opacities; these are then used to compute
the transfer of radiation through each slab, with the output used as input in the
next one. In such treatments, one typically introduces artificial turbulent
broadening of the resulting lines by $\sim$ 500 km/s through the parameter {\tt
vturb} of XSTAR. Note that no such broadening is necessary in our models; instead,
the lines are naturally broadened by the velocity shear of adjacent wind layers.
The profiles of {\sl all} lines shown in Figure 3 were computed by considering the
continuous absorption of radiation by each ion as its ionic fraction and velocity
vary with $r$ along the observer's LoS. As shown in Figure 4 the profiles begin
shallow, broad and highly blue-shifted at the smallest radii where a given ion is
formed. As $r$ increases, larger ionic fraction and lower velocity make the lines
less blue-shifted and deeper, to achieve their final shape at distances where $\xi$
becomes too small to support that ion. One should finally note that these
calculations employ only the radial component of the wind velocity, which depends
on the observer's inclination angle$^{14}$.

%The property that distinguishes the MHD from the radiation and thermally driven winds is
%their radial density profile $n(r) \propto 1/r^{1+\alpha}, \, (\alpha \ll 1)$, with $r$
%the radial coordinate along the observer's line of sight (LoS). This particular density
%profile implies for the ionization parameter $\xi \propto L/n(r) r^2 \propto
%r^{-(1+\alpha)}$ while for the wind column along LoS $N_H \propto
%\xi^{\alpha/(1-\alpha)}$

The broader behavior of our models, i.e. the relations among ($N_H,\xi,V$) where $N_H
\propto \xi^{\alpha/(1-\alpha)}, N_H \propto V^{2\alpha}$ and $V \propto \xi^{1/2(1 -
\alpha)}$, are a result of  their self-similarity (compare with $N_H \propto \xi^{-3}$
of a specific thermal wind model$^{23}$). The {\em Chandra} grating observations of GRO
J1655-40 (and also those of AGN) indicate a weak dependence of $N_H$ on $\xi$ ($\alpha
\simeq 0.2 \ll 1$). This behavior is consistent with observations: the high $\xi$
transitions of Fe, Ni and Co, are systematically broader and of higher velocity, of
order $V \sim 100 - 1,000$ km~s$^{-1}$ (implying $r \lsim 10^4 R_S$), with $N_H \sim 10^{23}$
cm$^{-2}$. The lower $\xi$ X-ray absorbers, on the other hand, are narrower and slower
($V \sim 100$ km~s$^{-1}$), implying larger distance from the black hole, with lower
$N_H (\sim 10^{22}$ cm$^{-2}$).
{ Detailed best-fit wind properties are listed in Table~S2 in Methods. }

As discussed elsewhere$^{20}$, the ionization  structure  of the wind plasma is
independent of the compact object mass $M$, once radii are scaled by $R_S$ and the mass
flux by $\dot M_{\rm Edd}$. For a given (normalized) mass flux $\dot m$, differences in
the wind properties are reduced to different values of the inclination angle $\theta$
and ionizing SED. Progressively increasing (decreasing) the X-ray
contribution of the SED leads to decreasing (increasing) absorber's velocities$^{6}$. As
such, the $V \simeq 100 - 1,000$ km/s of the H/He-like Fe absorbers of (the X-ray-dominated) GRO J1655-40, is considerably smaller than those ($V \sim 10^4$ km/s) of Seyfert
galaxies$^{3}$, whose SEDs are dominated by the big blue bump; in agreement with this
notion, BAL quasars, with the weakest X-ray contribution in their SED amongst AGN,
exhibit the highest absorber's velocities$^{6}$. Another issue associated with the
parameters of these winds is the presence of \fexxii\ lines at 11.75, 11.9 \AA, as these
are thought to be high density indicators$^{10}$; we have estimated that these lines
can be produced also by photon-excitation near the edge of our model winds, a fact
consistent with their observed velocities; this issue will be treated in more detail in
future work.
%Part of the low absorber velocities of GRO J1655-40 is attributed to the
%high inclination of this source (the wind velocity is mainly in the $\phi-$direction at
%$\theta \gsim 80^{\circ}$).
Lastly, another qualitative difference of the absorber's properties of GRO J1655-40 from
those of AGNs is the absence of low $\xi~ (\lsim 10^{2.5}$) absorbers in GRO J1655-40,
attributed to the spatial extent of its wind, limited by the size of the binary orbit
($r \gsim 10^{12.5}$ cm, $V \lsim 250$ km/s).

Finally, the radial density and velocity of these winds with radius, indicates that their (normalized) mass flux $\dot m \sim r^2 V(r) n(r) \propto r^{1/2 - \alpha}$,  is an
increasing function of $r$ for $\alpha < 1/2$. Therefore, most of the mass available for
accretion escapes before reaching the black hole. One should note that there is mounting
evidence that this is the case (wind mass flux similar to or higher than that needed to
power the continuum) both in AGNs$^{2}$ and in galactic X-ray binaries$^{29}$.

%The above model is sufficiently well defined and specific that it should be confirmed or
%refuted by additional observations and  analysis of galactic and extragalactic accretion
%powered X-ray  sources.

\begin{methods}

%{\bf METHODS}

%Here is a description of a specific method used.  Note that the subsection heading ends
%with a full stop (period) and that the command is \verb|\subsection{}| not
%\verb|\subsection*{}|.

\subsection{{\it Chandra}/HETGS Observation and Data Analysis.}

\gro\ was observed in its outburst phase at an exceptionally high cadence  with {\it
Chandra}/HETGS on 2005 April 1, starting at 12:41:44 (UT or MJD~53461.53) yielding a
total exposure of $44.6$ ks after the standard filtering.
We started with the same spectral data extracted by \cite{7,16} where a detailed data
reduction procedure is described and used subsequently as described in \cite{17} (and
references therein).
For subsequent analysis steps we used the {\tt heasoft v.6.16} package and the latest
calibration files.

%{\bf {\it Chandra}/HETGS data reduction.}
%We derived the Suzaku XIS cleaned event files and applied standard screening criteria. The 3?3 and 5?5 editing modes were combined. The source spectra were extracted from circular regions of 2.5ее radius centered on the source. The background spectra were extracted from annular regions with inner and outer radii of 3?4ее centered on the source and excluding contamination from the calibration sources. The spectra from the two front illuminated detectors, XIS0 and XIS3, were combined after verifying that the data are consistent with each other. Hereafter we refer to them as XIS03. The data of the back illuminated XIS1 detector are used as well.
%A detailed description of data reduction procedure is found in \cite{Miller08} and \cite{Kallman09}.
The standard gratings redistribution matrix file ({\tt rmfs}) was used to generate its
ancillary files ({\tt arfs}). We only used the first-order dispersive data in this
analysis. The net source count rate is $74.81$ cts~s$^{-1}$ and the observed flux is
$f_{\rm 2-10} =1.98 \times 10^{-8}$ ergs~cm$^{-2}$~s$^{-1}$ in $2-10$ keV.
%in units of counts s?1 in the E = 0.5?10 keV band are 0.0576, 0.0649 and 0.06 for the XIS0, XIS1 and XIS3, respectively. The background countrates are 18\% and 30\% of the XIS03 and XIS1, respectively.
%
All the spectra were background-subtracted and dead-time-corrected following the
standard procedures\cite{7,16}.

%{\bf {\it Chandra}/HETGS spectral analysis.} The spectral analysis was carried out based on the standard $\chi^2$-statistics using the software {\tt XSPEC v.12.8.2}. All uncertainties quoted are at $1\sigma$ level for one parameter of interest and wavelengths are reported in the observer's frame, unless otherwise stated.
%
Before applying our analysis, we examined the earlier X-ray studies \cite{7,16,17} in
order to securely establish the continuum component in the observed broad-band
spectrum. In agreement with the {\it RXTE} hard X-ray ($>2$ keV) spectrum, we adopted
the {\tt po} model to describe a power-law continuum with photon index $\Gamma=3.54$
(fixed) with an appropriate normalization $K_{\rm pl}$. We then employed multicolor
black body disk model ({\tt diskbb} in XSPEC) with maximum disk temperature $kT=1.34$
keV (fixed) and normalization of $K_{\rm diskbb}$ to match the estimated disk
luminosity. The latter component accounted for the dominant continuum flux. Therefore,
throughout our analysis we have employed spectral parameters identical with those used
in the literature. The ionizing luminosity in \gro\ was taken to be $L_{X} = 5.0 \times
10^{37}$ ergs~s$^{-1}$, in accordance with the  references\cite{7,16,17}, while the
disk inclination angle $\theta$ is constrained to be in the range $\ 67\deg - 85\deg$
\cite{7,30}.
%
%. The XIS03/XIS1 cross-normalization was left free to vary, but it was always found to be consistent within 3\%. We take into account the XIS/PIN cross-normalization of 1.16 Б 0.01.
%
All the spectra were adaptively binned to ensure an appropriate minimum of counts per
energy bin in order to perform the $\chi^2$-minimization in model fitting. Throughout,
we include the Galactic absorption due to neutral gas of hydrogen-equivalent column of
$N_H = 7.4 \times 10^{21}$ cm$^{-2}$ \cite{7,16,17,31}. This continuum spectral shape
was then fixed in the subsequent analysis of the X-ray absorbers to keep consistency.

\subsection{The MHD Accretion-Disk Wind Model.}

\noindent
Motivated by the exceptional S/N of the {\it Chandra}/HETGS data, we explored
their interpretation within the framework of global magnetohydrodynamic (MHD)
accretion-disk winds
%{\em ( condition of the ionized X-ray winds detected in \gro\ within a unique framework of  the well-defined MHD-driven accretion-disk wind models)}
which is primarily based on the well-defined MHD framework\cite{13}; having been applied
successfully to describe the major characteristics of AGN warm absorbers
(WAs)\cite{32,33,34,41} as well as the ultra-fast outflows (UFOs) of AGNs as diverse as
Seyfert galaxies\cite{3} and bright quasars\cite{35}. It is presently implemented to
model the ionized outflows associated with XRBs such as \gro.
%{\em ( by calculating a large-scale disk-wind launched by the action of a global magnetic field anchored to the disk surface)}.

To this end, we assumed different scalings of the magnetic flux with radius while
numerically solving the Grad-Shafranov equation within the ideal MHD
approximation\cite{5,6,18,19,20,36,37}. The radial profile of the wind density
considered in this study is given by $n(r) \propto r^{-(1+\alpha)}$; we have explored
four different cases, corresponding to  $\alpha=-0.1, 0, 0.2$ and $0.4$, each
representing a slightly different distinct global wind profile (see later description for details). We incorporate the wind structure into our radiative transfer
formulation with radiation spectrum given by the SED of \gro\ (dominated by the thermal
multicolor disk component, {\tt diskbb}, of its high/soft state during the observation)
and calculate its ionization equilibrium under heating-cooling balance in the wind
using the {\tt xstar} photoionization code (v.2.2.1bn21)\cite{28} in an approach
similar to our previous works\cite{5,6,19,20}. In the current wind model, we considered
different sets of parameters as listed in Table~S1. %; i.e. global density profile
%$\alpha=-0.1, 0, 0.2$ and $0.4$, inclination angle $\theta =70\deg$ and $80\deg$
%\cite{30}, and wind density normalization $n{\rm 17}=0.024-400$.
%
The basic global MHD wind structure is shown in Figure~S1: The poloidal magnetic field
lines (thick solid) are plotted along with the 2D density distribution $n(r,\theta)$
(in color), its contours (thin solid) and contours of the ionization parameter $\xi$
(dashed), as a function of cylindrical coordinates $(R,Z)$ in units of the innermost
launching radius $r_0$. The density $n(r,\theta)$ is normalized to its maximum value.
Note the $\textmd{linear}(Z)-\log(R)$ character of this Figure where the line-of-sight
(LoS) is not a straight line. Plasma in the wind is efficiently accelerated along each
field line from the disk surface which is treated as a boundary condition in this
study. The major acceleration phase takes place by the action of magnetic process
before the plasma becomes Alfv\'enic. The primary ionizing X-ray photons are produced
in the  innermost disk region which is treated as a point-like source in our radiative
transfer calculations.

Implementing a total of approximately 80 line transitions as well as edges from a series of commonly
observed elements (i.e. Ne, Mg, Si, S, Ar, Ca, Cr, Mn, Fe, Co, Ni) of various
ionization states, our model provides a unique realization of the observed WAs in \gro\
in a coherent context, in which each X-ray absorbing ion is explicitly associated with
a specific segment of the same contiguous disk-wind. In this approach, the
emergence/properties of the different X-ray absorbing ions, from the soft band
($\lambda \sim 12.1\aa$) to the Fe K band ($\lambda \sim 1.5\aa$), are not independent
of each other, but coupled and constrained by their common underlying wind structure
(as they ought to be), reducing the arbitrariness in our calculations. Therefore, the
proposed X-ray absorber model of \gro\ is not only internally consistent but also
strongly constrained globally.
%All line
%transitions from the soft band ($\lambda \sim 12.1\aa$) to the Fe K band ($\lambda \sim
%1.5\aa$) are coupled to each other, thereby eliminating arbitrariness in our
%calculations.
%
%In other words, modeled WAs in the broad range of $1$ keV $\lsim E \lsim $ $9$ keV in this work  are uniquely determined by the model leaving very little arbitrary flexibility in spectral calculations. Hence, the characteristic differences of WA properties among various ions (e.g. ionization parameter, column, velocity and so on) are simply attributed to their response to the irradiating SED within a continuous disk-wind driven by the MHD process.

%We will briefly describe the generic feature of our MHD-driven winds in \S 2 along with our computation for constructing a grid of simulated line spectra for subsequent data analysis. In \S 3 we show our fitting results  based on a 45-ks {\it Chandra}/HETG spectrum of \gro\
%making comparison between data and the models of thre different density profiles representing a different global behavior of the wind.
%%
%%deriving the best-fit values for the primary model variables.
%%
%We summarize and discuss the implications of the model in \S 4.

%
Our model is scale-invariant, thus the entire 2D wind structure is independent of the
BH mass for a given set of characteristic wind variables. These are\cite{13,19} the
particle-to-magnetic flux ratio $F_0$, the angular momentum $H_o$ and the initial
launching angle $\theta_0$. These variables uniquely determine, accordingly, the rest
of the dependent parameters, such as the rotation of the magnetic field lines
$\Omega_0$ and the specific Bernoulli function $J_0$\cite{5,13,19}.
{We choose $F_0 = 0.065, H_0 = -1.7, \Omega_0 = 1.016$ and $J_0 = -1.516$ in our
fiducial model\cite{5,13}.}
Our computational method applies the same approach used in our previous and other
works\cite{5,36,37}; i.e. we assume a geometrically-thin accretion disk at the equator
($\theta=\pi/2$) as a boundary condition where accreting plasma is in the Keplerian
motion (i.e. $V_\phi \sim V_{\rm K} \propto r^{-1/2}$). Due to the poloidal field
component and compression of the toroidal field,  plasma is magneto-centrifugally
launched by the $\mathJ \times \mathB$ force from the disk surface with $\mathv =(V^r,
V^\theta, V^\phi)=(0, 0.01, V_{\rm K})$ and it is efficiently accelerated along a
streamline up to a few times the initial Keplerian velocity at large
distances\cite{13,19}.
With the initial conditions on the disk (denoted by the subscript ``0") at $(r = r_0
\simeq r_{\rm ISCO}, \theta = \theta_0 = \pi/2)$, the density normalization at $(r_0,
\theta_0)$ around a BH of mass $M$ is expressed as
\begin{eqnarray}
n_0 \equiv \frac{f_w \dot{m}_a} {\sigma_T R_S}  = \tilde{n}_{17}(f_w, \dot{m})
\left(\frac{M}{10\Msun}\right)^{-1}  ~\textmd{cm$^{-3}$} \ ,
%
%\simeq  2.5 \times 10^{17} \left(\frac{\dot{m}_a}{0.5}\right) \left(\frac{M}{10\Msun}\right)^{-1}  ~\textmd{cm$^{-3}$}  \ ,
%
\label{eq:no}
\end{eqnarray}
where $\sigma_T$ is the Thomson cross-section, $\dot{m}_a$ is the dimensionless
mass-accretion rate, $\tilde{n}_{17}$ is the wind density factor in units of
$10^{17}$ cm$^{-3}$, and $f_w$ is the ratio of the outflow rate in the wind to
accretion at $r = r_0 \simeq r_{\rm ISCO}$ for $M=7\Msun$ BH mass relevant for
\gro. The exact values for these quantities, however, depend on the specific source
and its specific spectral state. In our model, we let $\tilde{n}_{17}$  vary also,
among others, to search for the best-fit spectrum as found in Figures~3 and 4.
Then, a global density distribution on $(r,\theta)$-plane is determined by
\begin{eqnarray}
n(r,\theta) = n_0 g(r) f(\theta) \ ,
\end{eqnarray}
where the angular dependence $f(\theta)$ is numerically solved by the
Grad-Shafranov equation (the momentum-balance equation in a direction perpendicular
to a  field line) in an ideal MHD framework. For the observed $L_{\rm ion} = 5.0
\times 10^{37}$ ergs~s$^{-1}$ \cite{9,10,11}, one can estimate that
\begin{eqnarray}
\xi(r,\theta) \equiv \frac{L_{\rm ion}}{n(r,\theta) r^2}
%= 1.4 \times 10^8
~~ \textmd{erg~cm~s$^{-1}$}  \ ,
\end{eqnarray}
for a given set of ($\tilde{n}_{17}, \theta, \alpha$) as listed in Table~S1.

\subsection{X-ray Photoionization of Disk-Winds.}

The definition of the photoionization parameter $\xi$, which determines the wind plasma
ionization, involves the local radiation field, which is in general different from that
of the input SED because of absorption along the wind. The correct study of wind
ionization therefore requires that the input SED be transferred correctly to each wind
point numerically.
To this end, we performed radiative transfer calculations of the input X-ray spectrum employing {\tt xstar} %(KallmanBautista01; v2.2.1bn21)
to compute the the local plasma ionization and resulting opacity as a function of the
wavelength along a given LoS.
Using the ionizing spectrum of the baseline continuum  of \gro\ described above, we
discretize the wind along a given LoS of angle $\theta$ into a large number of
contiguous zones of logarithmically constant width $\Delta r/r \simeq 0.15$,
sufficiently small to consider each radial zone as plane parallel\cite{5}.
%
%
%We adopt the same computational scheme as used in our previous calculations (FKCB10a, FKCB10b, F15) in order to assess local ionization equilibrium of the irradiated plasma along the LoS; i.e.
%local radiative transfer calculations are carried out from one zone to the next in a progressive manner including the local opacity and emissivity at each radius for many major ions based on the wind kinematics from the ideal MHD calculations. Note that the model naturally provides an inherent velocity gradient (i.e. shear velocity) along a LoS which would otherwise be invoked arbitrarily as {\it turbulent velocity} in the other models.
%
Having obtained the ionic column $N_{\rm ion}$ appropriate for each line transition from
the photoionization calculations  one can compute line spectra through the local line
optical depth
\begin{eqnarray}
\tau_\nu(r,\theta) \equiv \sigma_{\rm photo,\nu} N_{\rm ion} \ ,
\end{eqnarray}
where $\sigma_{\rm photo,\nu}$ is the line photoabsorption cross-section at frequency
$\nu$ calculated by assuming a Voigt profile for the transition with the local wind
shear $dV/dr$ for line broadening in place of the plasma turbulent velocity\cite{38}
and integrating the local absorption over all distances along the LoS.

In the commonly employed phenomenological fitting analyses, the absorbers across the
wavelength range are typically treated as mutually-independent components arising from
physically-disconnected regions. In contrast, the proposed model has to and does
account for all the absorbers simultaneously  from the same continuous global
disk-wind.%%
%The current  model is physically motivated by the well-defined MHD process that
%can provide a broad-band absorption features across the entire X-ray band in
%a self-consistent fashion. In contrast to the conventional phenomenological fitting approach,
%
%describe the universal magnetic structure of black hole accretion disks in a global
%perspective.

In the end, we implement our MHD-wind model, \verb"mhdabs", into \verb"xspec" as a
multiplicative table model. The symbolic spectral form reads as
\begin{eqnarray}
{\tt tbabs*(po+diskbb)*mtable\{mhdabs\}} \ ,
\end{eqnarray}
where we have used the previously estimated values of parameters such as the Galactic
absorption due to neutral hydrogen column ({\tt tbabs}) is $N_H = 7.4 \times 10^{21}$
cm$^{-2}$ \cite{31} and the black hole mass of $M=7\Msun$.

%\subsection{Physical Conditions of Magnetically-Driven X-ray Absorbers.}
\subsection{Physical Parameters of Magnetically-Driven X-ray Absorbers.}

The present study makes a straightforward, yet physically motivated, approach in
modeling the  properties of X-ray absorbers of \gro\ within the framework of a well
defined class (MHD) of accretion disk wind models. Figure~3 shows the $46$ ks {\it
Chandra}/HETG broadband spectrum of \gro\ between $\sim 1.5\aa - 12 \aa$ with the
best-fit {\tt mhdabs} model of $\alpha=0.2$ and $\tilde{n}_{17}=9.3$; in all our
calculations we assume solar ion abundances ($A_{\rm ion}=1$) for all elements, as the
simplest, less contrived assumption. In producing Figure 3 as well as all similar
figures we let the wind density normalization $n_0$ (thus $\tilde{n}_{17}$) vary
along with the value for $\alpha$ in a way that the model produces the correct value of
most ionic columns. The disk inclination angle $\theta$ was also allowed to vary within
the narrow angle range allowed by observation and was set to $80 \deg$ in all figures.
{ We present the dependence of individual lines on the wind density normalization
$\tilde{n}_{\rm 17}$ for $\alpha=0.2$ in Figure~S2. It is demonstrated that the
global spectral fit at longer (\fexxiii\ near $\sim 11\aa$) and shorter (\caxix,
\caxx\ and \arxviii\ near $3.1\aa$) wavelengths clearly favors $\tilde{n}_{\rm 17}
= 9.3$.}
{ Detailed absorber's properties are listed in Table~S2. }
We see that the model captures correctly most of the features of the spectrum.
Discrepancies in a few transitions could be attributed either in different abundances or
in the fact that the wind density might deviate locally from the smooth, contiguous
power-law dependence assumed throughout this work.

{ Besides the spectra for different values of $\alpha$  in Figure~S3, we also
present in Figure~S4 the ratios of the observed to modeled line equivalent widths
(EW) for major ions, as done in other related work\cite{7,23} along with their mean
(solid green) and median (dotted green) values also indicated by numbers.
%We have computed the average value of the sum of absolute values
%of the ratios of these differences from 1,
We have further calculated the mean values of their relative deviations
$\Delta_{\rm EW} =|$(EW(obs) $-$ EW(model))$|/$ EW(model) and we found the following pairs of values
$(\alpha, \Delta_{\rm EW})=(-0.1, 0.54), (0.0, 0.44), (0.2, 0.24), (0.4, 0.47)$.
The ratios here serve as a measure of the goodness of the global fit for our
calculations. With this measure we have demonstrated that the $\alpha=0.2$ wind
(with the best-fit value of $\tilde{n}_{\rm 17}=9.3$) shown in Figure~3 is indeed
statistically most favored over the other wind structure in support for our
conclusion. }
%($\Delta(\rm EW) = 0.54, 0.44, 0.24, 0.47$, respectively, for $\alpha
%=-0.1,0.0,0.2,0.4)$ with the value $\alpha =0.2$ preferred with this statistical
%measure.

\vspace{0.5cm}

\subsection{Mass and Energy Budget of X-ray Winds in \gro.}

The MHD winds that are launched from an accretion disk as discussed above, can
carry power \textbf{comparable to or larger than their radiant one} and mass flux
\textbf{larger than that needed to power their luminosity by accretion}. These can
have significant influence on their surroundings, especially in the AGNs that are
located at the centers of galaxies or clusters. These winds are also of interest in
XRBs for determining the energy budget of the system.

One generally estimates the mass flux and luminosity associated with a specific
transition of a known $\xi-$value and measured velocity $V$. Since
\begin{eqnarray}
\dot{M}_{\rm out}^{\rm (local)} \simeq n(r) r^2 V  ~~~
{\rm and} ~~~ \xi = \frac{L_X}{n(r) r^2} ~~~{\rm then} ~~~
\dot{M}_{\rm out}^{\rm (local)} \simeq \frac{L_X}{\xi}V \, ,
\end{eqnarray}
a general relation, applicable to any transition of known $\xi$ and measured
$V$. For winds with density scaling given by our model, $\xi$ is
proportional to $r^{-(1+\alpha)}$ and $V \propto r^{-1/2}$ leading to
\begin{eqnarray}
\dot{M}_{\rm out}^{\rm (local)} \simeq n(r) r^2 V \propto r^{1/2-\alpha}  ~~~
\end{eqnarray}
indicating that for $\alpha<1/2$ the mass flux in the wind increases with
distance\cite{8,13}. So, with the above scalings, while there is mass flux of fully
ionized plasma (not discernible in the spectrum) at radii smaller than those at
which the Fe features occur (namely at $ r \gsim 10^4 R_S$), this mass flux is a
small fraction of that associated with the partially ionized plasma. For the value
of $\tilde{n}_{\rm 17}$ appropriate to our best fit of the data, we obtain $\dot M
\sim 10^{17} (r/R_S)^{1/2-\alpha}$ g s$^{-1}$ whose value, for $r \sim R_S$, is
comparable to that needed to produce the X-ray luminosity of this source. \textbf{
This is consistent with a value $f_w \sim 1$ for the parameter of equation~(1) that
determines the ratio of the accreted to wind mass flux near the disk inner edge, in
that it reproduces the observed radiation with efficiency of $\simeq 0.1$. }

Finally, one can further estimate the corresponding kinetic power due to mass-loss since
the kinetic power scales as
\begin{eqnarray}
\dot{E}_{\rm out}^{(\rm local)} \propto \dot{M}_{\rm out}^{\rm (local)} v_{\rm out}^2
\sim r^{-(\alpha+1/2)} \ .
\end{eqnarray}
This expression indicates that, for the scalings of our model, even though most
mass loss comes from the largest wind radii, most of the wind kinetic energy comes
from the wind segments launched in the black hole proximity. { With the best-fit
value of $n_0$ derived from our analysis, the wind kinetic luminosity is on the
order of $\dot{E}_{\rm kin} \lsim 10^{38}$ erg s$^{-1}$, comparable with its photon
luminosity and broadly consistent with the fact that for $r \sim R_S$ the mass flux
in the wind is comparable to that of accretion. Considering that the Fe transitions
are formed at radii $r \gsim 10^4 R_S$, the kinetic luminosity attributed to the Fe
emitting plasma is smaller by the factor of $R_S/r \sim 10^{-4}$ or $\dot{E}_{\rm
kin, Fe} \sim 10^{33}$ erg s$^{-1}$. } \textbf{One should note that for a given
$\dot m$ all winds carry the same kinetic power per unit black hole mass, $\dot
E_{\rm out}/M$. However, because of their higher ionization and formation of the \fexxv/\fexxvi\ absorbers at larger (normalized) distances, the XRB's $\dot E_{\rm out}/M$ will
be smaller than those of AGN, as documented to be the case in a recent work$^{42}$.}

%\subsection{Global Nature of Magnetically-Driven X-ray Absorbers.}
\subsection{Assessment of the Different Wind Density Profiles.} Figure~S3 gives the
broad-band spectra of our models for the different values of $\alpha$ {for which
$\tilde{n}_{\rm 17}$ is constrained by the broad-band fitting. }
%(in units of $10^{17} \;{\rm cm}^{-3}$).
As discussed earlier, successful models should provide the correct ($N_H, V$) at
the values of $\xi$ that produce a given ionic species. Clearly, such a fit is by
necessity statistical given the broad range of $\xi$ at which the various ions of
the spectrum appears. Our density parametrization allows a global view of this
notion. The optimal value of $\alpha$ was chosen so that it minimizes the
differences between observed and computed EW of the ensemble of the transitions by
varying $\tilde{n}_{\rm 17}$ (see Fig.~S4) in the fitting procedure. {Changing
$\tilde{n}_{\rm 17}$ for a given $\alpha$ thus minimizes deviations of calculation
from observation at the band associated with given value of $\xi$, to which the
particular choice is tailored to. See also Figure~S2 for the significance of
$\tilde{n}_{\rm 17}$.} However, the wrong value of $\alpha$, then, produces
increasingly large deviations from observations either at lower or larger $\xi$.
So, we see that for $\alpha = -0.1$ (green) we underestimate the EW of short
wavelength transitions while for $\alpha=0.4$ (orange) the low ionization
transitions have very small columns, excluding this value off-hand.
{Thus, Figures~S3 and S4 combined together clearly indicates that
%
%Figure~S4 provides the ratios of the observed to model
%EW for the most significant transitions, along with their average (black) and median
%values (gray).
%
the model with $\alpha = 0.2$ is preferred by these metrics at a statistically
significant level.}

Clearly, better fits to the data can be achieved by judiciously introducing breaks in
the wind density profile to provide better fits to the local plasma column. Such breaks
are not excluded by any underlying principle but only by the self-similarity
requirement of our solutions. However, we have chosen not to do so. Such an approach
might be the subject of future investigations.

\textbf{In closing, we would like to draw attention to another recent work on XRB MHD-winds 
%Chakravorty et al.
$^{40}$ studying the photoionization of the accretion disk winds of the  similar MHD-driven model
%Ferreira
$^{15}$ as they relate to the absorbers of \gro. These winds have steeper
density profiles ($\alpha \gsim 0.4$) than those employed herein and in our earlier AGN
work; as such, the values of $\xi$ that allow for the presence of \fexxvi\ (used as a measure of the wind ionization) occur at larger distances than in the
winds employed herein. Given the steeper density profiles, the corresponding
columns are smaller than observation implies (see our Figs.~S3 and S4), while in the absence of
detailed absorption line modeling it is uncertain how well that model would fit
the data. On the other hand, their wind models relate the density and magnetic flux
distribution closer to the dynamics of the accretion disk than those of our model 
through a balance between field accretion and diffusion. We plan to address this
issue, as it relates to the winds we have so far employed, in a future work.
}

\end{methods}

%\bigskip

{\bf References}

%\bigskip

%\end{methods}

%% Here is the endmatter stuff: Supplementary Info, etc.
%% Use \item's to separate, default label is "Acknowledgements"

\begin{addendum}

\item We are grateful to Tim Kallman for providing us with the {\it Chandra}/HETG data for \gro. KF, DK and CS acknowledge support by a NASA/ADP grant. EB received funding from the European Unions Horizon 2020 research and innovation programme under the Marie Sklodowska-Curie grant agreement no. 655324. and from the I-CORE program of the Planning and Budgeting Committee (grant number 1937/12). 
Support for this work was in part provided by the National Aeronautics and Space Administration through Chandra Award Number AR6-17013A issued by the Chandra X-ray Observatory Center, which is operated by the Smithsonian Astrophysical Observatory for and on behalf of the National Aeronautics Space Administration under contract NAS8-03060.

\item[Competing Interests] The authors declare that they have no
competing financial interests.

\item[Correspondence] Correspondence and requests for materials
should be addressed to K. Fukumura~(email: fukumukx@jmu.edu).

\end{addendum}

%%
%% TABLES
%%
%% If there are any tables, put them here.
%%

\newpage

\begin{figure} %=========== fig.1
\begin{center}
\includegraphics[trim=0in 0in 0in
0in,keepaspectratio=false,width=4.0in,angle=0,clip=false]{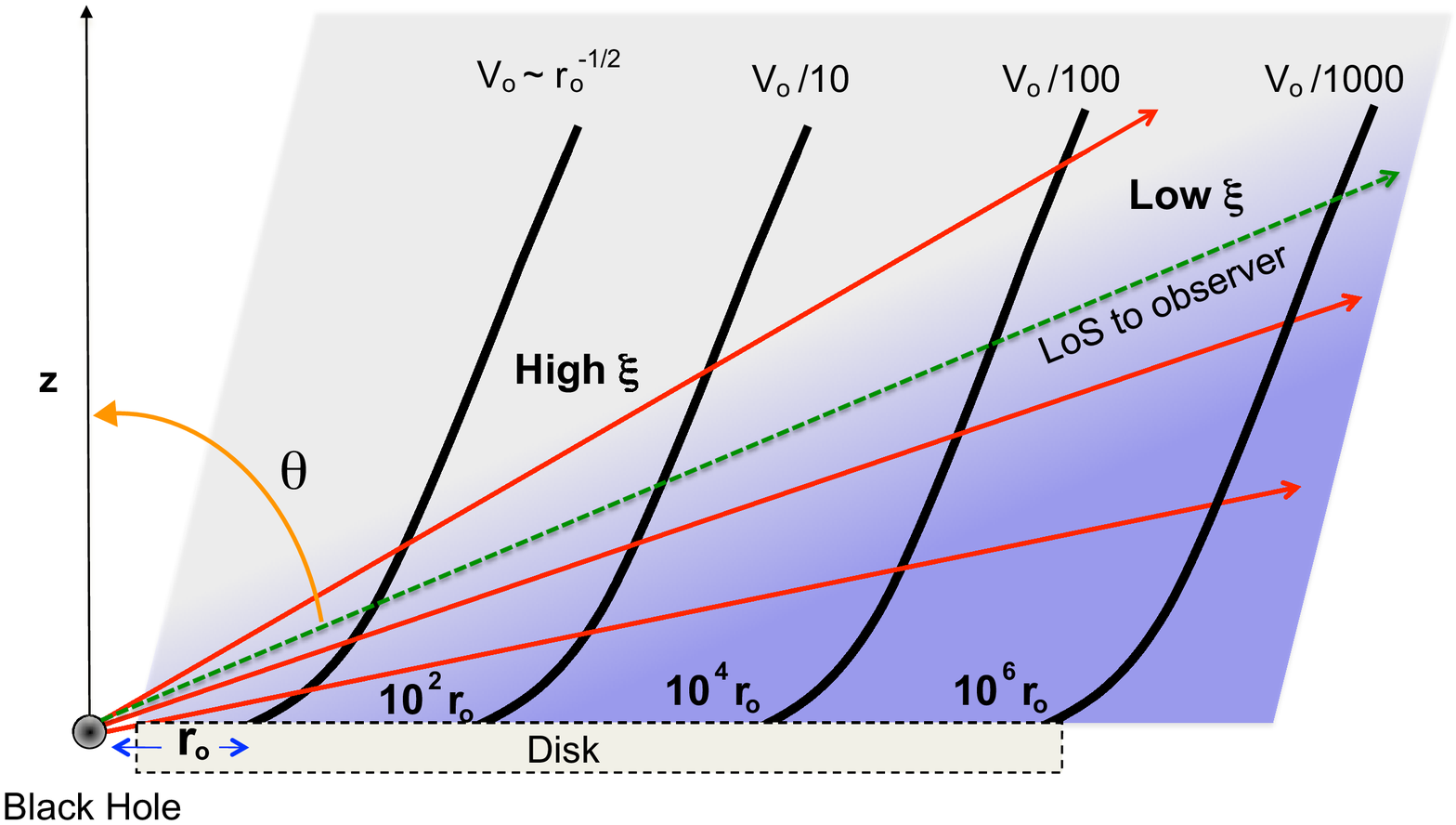}
%{poloidal.eps}
\caption{{\bf Schematic of an MHD accretion disk wind.} Poloidal 2D wind streamlines (thick solid), the decreasing  velocity ($v_0 \approx c$) and ionization ($\xi$) with radius is illustrated. The hatched region represents the absorber of the \gro\  wind region with 100-1,000 km\,s$^{-1}$. The arrows indicate possible line-of-sight (LoS) effects with the green arrow believed to be the true LoS based on published binary solutions. }
%\caption{\footnotesize {The Poloidal 2D wind streamlines (thick solid) along with the density distribution $n(r,\theta)$ (in color) and contours (thin solid) and contours of the ionization parameter $\log \xi$ (dashed), as a function of cylindrical distances $(R,Z)$ in units of the innermost launching radius $R_0$. Note the linear (Z) -- log(R) character of the figure, where the LoS is not a straight line.}}
\end{center}
\end{figure}

\vskip 10pt

{\sf \noindent \textbf{Figure 1}
Schematic of an MHD accretion disk wind.

\noindent Poloidal 2D wind streamlines (thick solid), the decreasing  velocity ($v_0 \approx c$) and ionization ($\xi$) with radius is illustrated. The hatched region represents the absorber of the \gro\  wind region with 100-1,000 km\,s$^{-1}$. The arrows indicate possible line-of-sight (LoS) effects with the green arrow believed to be the true LoS based on published binary solutions.}
%The Poloidal 2D wind streamlines (thick solid) along with the density distribution $n(r,\theta)$ (in color) and contours (thin solid) and contours of the ionization parameter $\log \xi$ (dashed), as a function of cylindrical distances $(R,Z)$ in units of the innermost launching radius $R_0$. Note the linear (Z) -- log (R) character of the figure, where the LoS is not a straight line. }

\newpage

\begin{figure} %=========== fig.2
\begin{center}
\includegraphics[trim=0in 0in 0in
0in,keepaspectratio=false,width=4.0in,angle=0,clip=false]{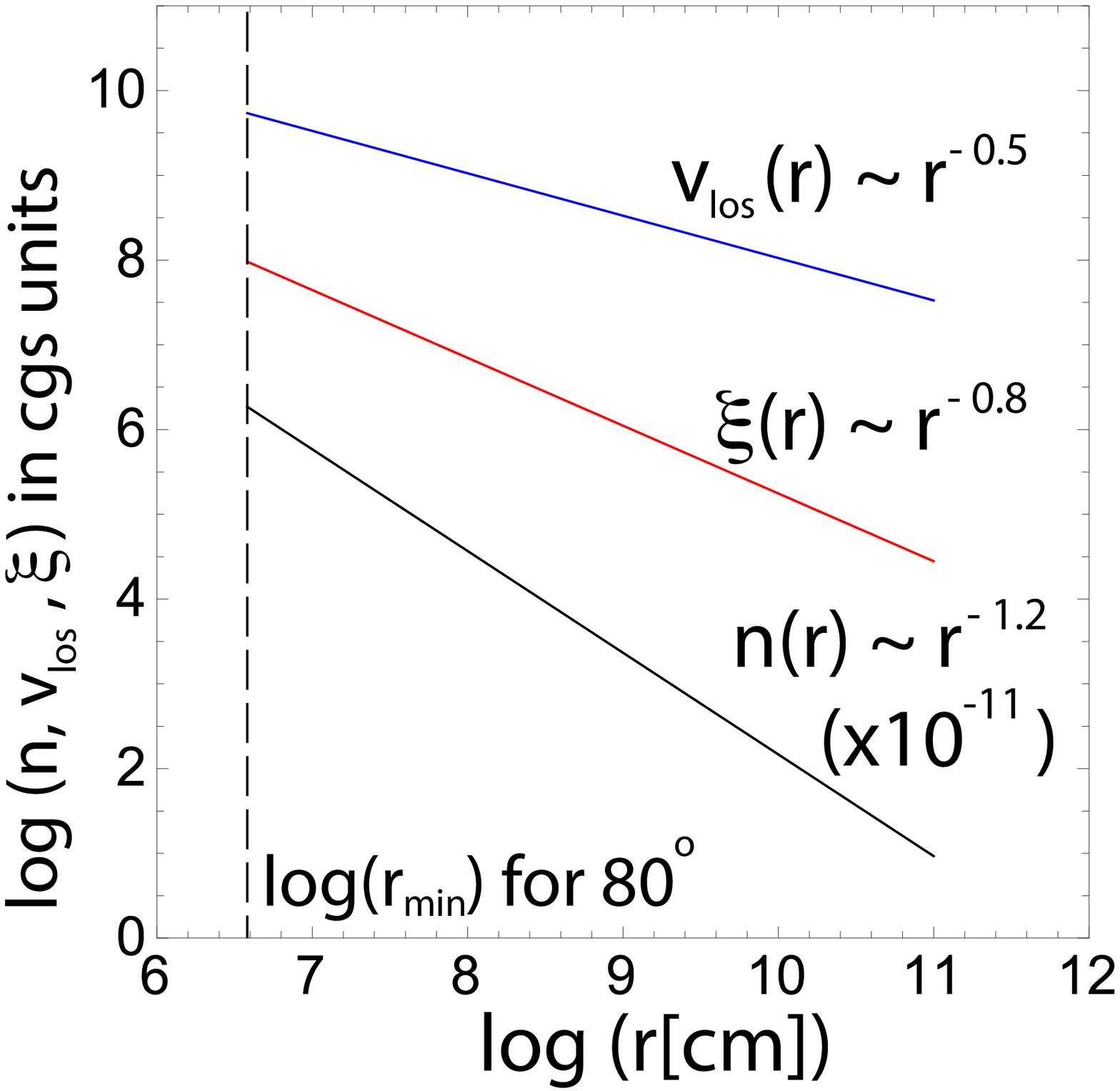}
\caption{{\bf Radial wind profiles inferred from our best-fit model of $\alpha=0.2$.} Wind density $n$[cm$^{-3}$] (black), the
LoS wind velocity $V$ (cm/s) (blue) and the ionization parameter $\xi$ [erg~cm~s$^{-1}$] (red)
for $\theta_{\rm obs} = 80 \deg$. The vertical line indicates the location of the innermost
wind streamline for $80\deg$. Note that $\log(n)$ is offset by -11.0 in the vertical
direction for presentation purposes.}
\end{center}
\end{figure}

\vskip 10pt

{\sf \noindent \textbf{Figure 2}
Radial wind profiles inferred from our best-fit model of $\alpha=0.2$.

\noindent Wind density $n$[cm$^{-3}$] (black), the
LoS wind velocity $V$ (cm/s) (blue) and the ionization parameter $\xi$ [erg~cm~s$^{-1}$] (red)
for $\theta_{\rm obs} = 80 \deg$ and $\alpha=0.2$. The vertical line indicates the location of the innermost
wind streamline for $80\deg$. Note that $\log(n)$ is offset by -11.0 in the vertical
direction for presentation purposes.}

\newpage

\begin{figure} %=========== fig.3
\begin{center}
\includegraphics[trim=0in 0in 0in
0in,keepaspectratio=false,width=4.99in,angle=-0,clip=false]{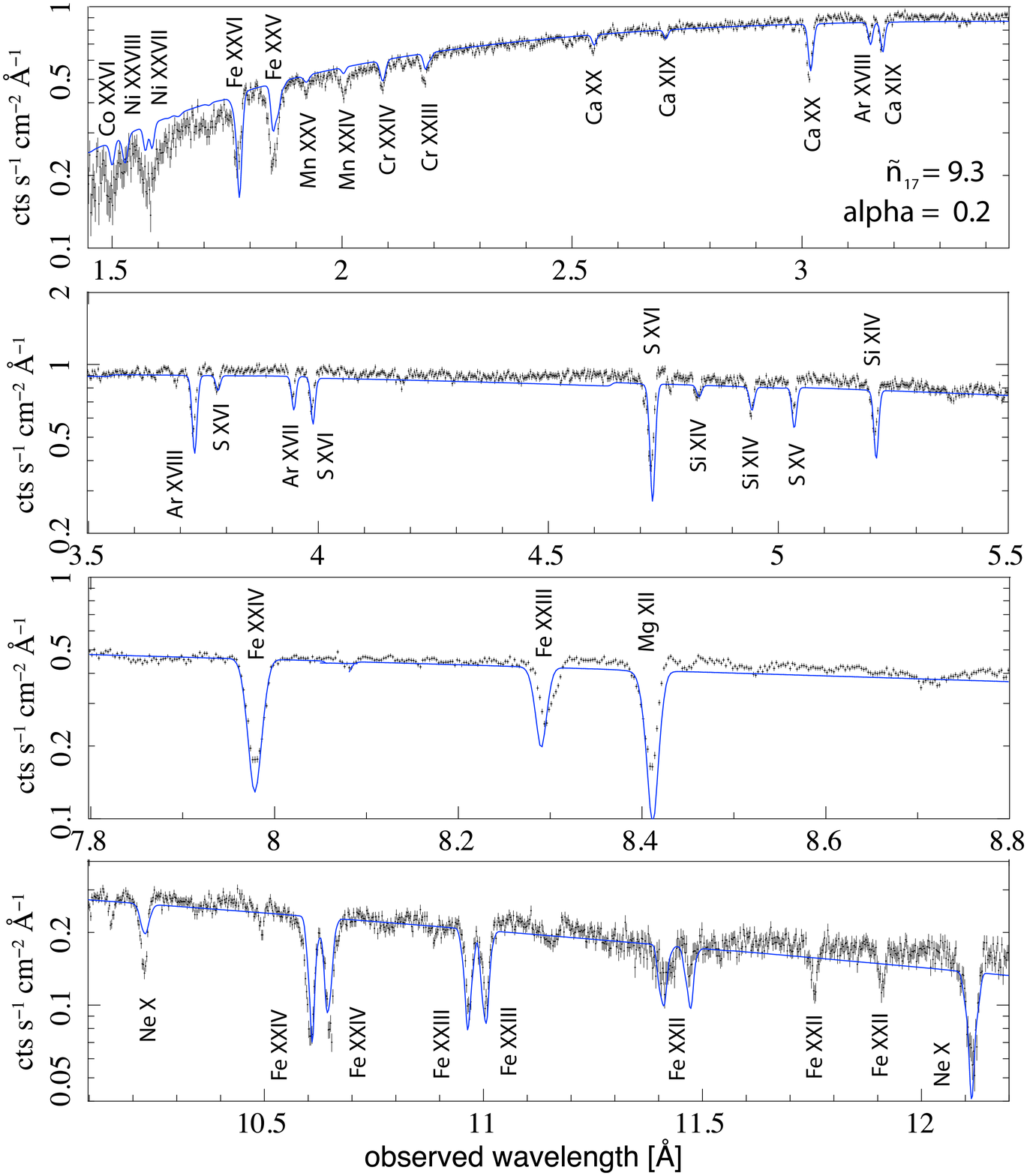}
%{broadband5.eps}
\caption{{\bf The 46-ks {\it Chandra}/HETG spectrum of \gro\
overlaid on the global MHD-wind model for $\alpha = 0.2$ (blue line) with $\tilde{n}_{\rm 17}=9.3$.} Adopting the previously constrained baseline continuum (i.e. a multicolor
disk model {\tt diskbb} of $kT=1.35$ keV with $L_x = 5 \times 10^{37}$ erg~s$^{-1}$), the composite line spectra are computed assuming solar abundances for all ions as discussed in the text and shown in detail in Fig.~4. We find $\theta=80\deg$ and $\tilde{n}_{\rm 17} = 9.3$ at $r_o = 3.81 \times 10^6$ cm with $M=7\Msun$ black hole for the best-fit spectrum. The best-fit wind parameters are listed in Table~2 in Supplementary Materials. }
\end{center}

\end{figure}

\vskip 10pt

{\sf \noindent \textbf{Figure 3}
The 46-ks {\it Chandra}/HETG spectrum of \gro\
overlaid on the global MHD-wind model for $\alpha = 0.2$ (blue line) with $\tilde{n}_{\rm 17}=9.3$.

\noindent
Adopting the previously constrained baseline continuum (i.e. a multicolor
disk model {\tt diskbb} of $kT=1.35$ keV with $L_x = 5 \times 10^{37}$ erg~s$^{-1}$), the composite line spectra are computed assuming solar abundances for all ions as discussed in the text and shown in detail in Fig.~4. We find $\theta=80\deg$ and $n_o = 9.3 \times 10^{17}$ cm$^{-3}$ at $r_o = 3.81 \times 10^6$ cm with $M=7\Msun$ black hole for the best-fit spectrum. The best-fit wind parameters are listed in Table~2 in Supplementary Materials.
}

\newpage

\begin{figure} %=========== fig.4
\begin{center}
%\hskip -42pt
\includegraphics[trim=0.0in 0in 0in
0in,keepaspectratio=false,width=3.0in,angle=-0,clip=false]{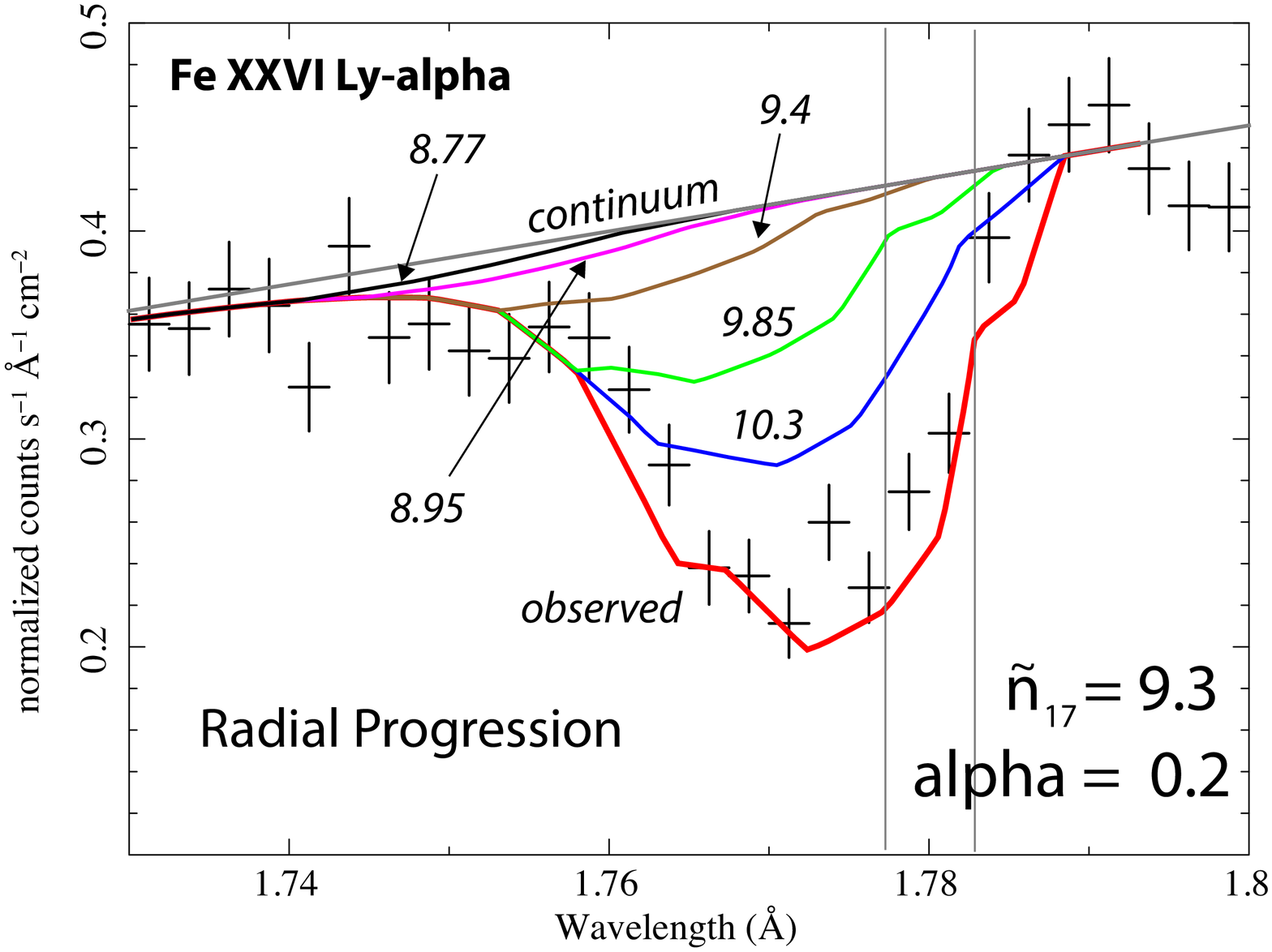}
\includegraphics[trim=0.0in 0.6in 0in
0in,keepaspectratio=false,width=3.4in,angle=-0,clip=false]{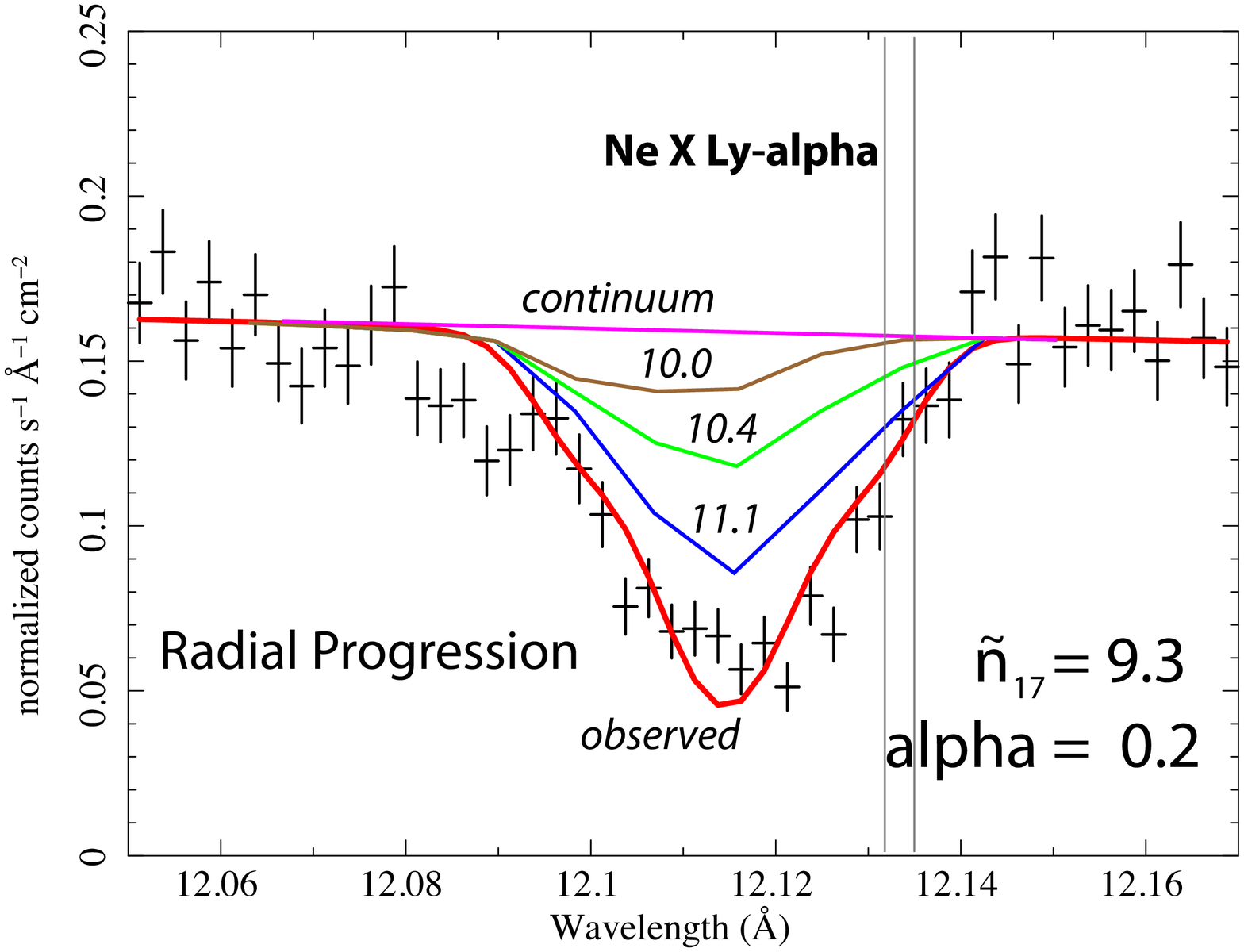}
\caption{{\bf Modeling the \fexxvi\ Ly$\alpha$ doublet (1.778, 17834 \AA) and the \nex\ Ly$\alpha$ doublet (12.132, 12.136 \AA) with the MHD winds of $\alpha=0.2$ and $\tilde{n}_{\rm 17}=9.3$.} The progressive development of each  spectrum with increasing distance $r$ is shown by the sequence of colored lines convolved with the {\it Chandra}/HETG resolution and overlaid on the data (the red line corresponds to
the line profile as seen at infinity). The profiles were computed employing the values
of ($n_0, \alpha, \theta$) of Figure~3. The double vertical lines denote the rest frame wavelength of these transitions, while the numbers at each colored line denotes the value of $\log r\textmd{[cm]}$ along the LoS. One should note that the model spectra (in color) are not fit to the data but an {\it a priori} calculation overlaid on \gro\ data. }
\end{center}
\end{figure}

\vskip 1pt

{\sf \noindent \textbf{Figure 4}
Modeling the \fexxvi\ Ly$\alpha$ doublet (1.778, 17834 \AA) and the \nex\ Ly$\alpha$ doublet (12.132, 12.136 \AA) with the MHD winds of $\alpha=0.2$ and $\tilde{n}_{\rm 17}=9.3$.

\noindent The progressive development of each  spectrum with increasing distance $r$ is shown by the sequence of colored lines convolved with the {\it Chandra}/HETG resolution and overlaid on the data (the red line corresponds to
the line profile as seen at infinity). The profiles were computed employing the values
of ($n_0, \alpha, \theta$) of Figure~3. The double vertical lines denote the rest frame wavelength of these transitions, while the numbers at each colored line denotes the value of $\log r\textmd{[cm]}$ along the LoS. One should note that the model spectra (in color) are not fit to the data but an {\it a priori} calculation overlaid on \gro\ data.  }

% ============== figures for Methods section

\begin{figure} %=========== S.1
\begin{center}
\includegraphics[trim=0in 0in 0in
0in,keepaspectratio=false,width=4.0in,angle=0,clip=false]{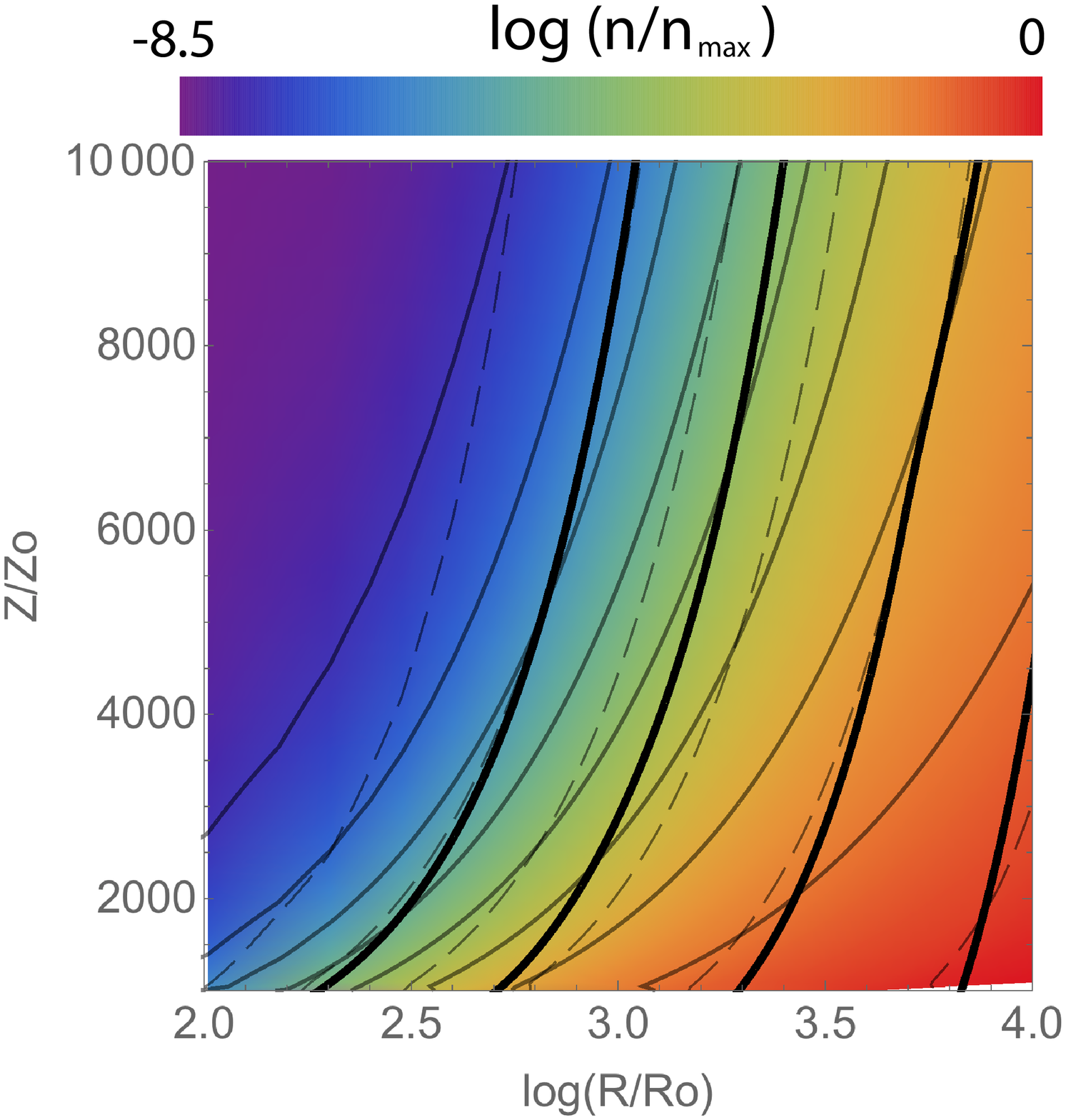}
\caption{{\bf Fig.~S1. \\ A poloidal 2D view of the fiducial MHD-driven disk-wind.} .
We show the magnetic field lines (thick solid) along with the density distribution $n(r,\theta)$ (in color) and contours (thin solid) and contours of the ionization parameter $\log \xi$ (dashed), as a function of cylindrical distances $(R,Z)$ in units of the innermost launching radius $R_0$. Note the $\textmd{linear}(Z)-\log(R)$ character of this Figure, where the line-of-sight (LoS) is not a straight line. }
\end{center}
\end{figure}

\vskip 0pt

{\sf \noindent \textbf{Figure S1.}
A poloidal 2D view of the fiducial MHD-driven disk-wind.

\noindent We show the magnetic field lines (thick solid) along with the
density distribution $n(r,\theta)$ (in color) and contours (thin solid) and contours of
the ionization parameter $\log \xi$ (dashed), as a function of cylindrical distances
$(R,Z)$ in units of the innermost launching radius $R_0$. Note the
$\textmd{linear}(Z)-\log(R)$ character of this Figure, where the line-of-sight (LoS) is
not a straight line.}

\clearpage

\begin{figure} %=========== S.2
\begin{center}
\includegraphics[trim=0in 0in 0in
0in,keepaspectratio=false,width=3.2in,angle=0,clip=false]{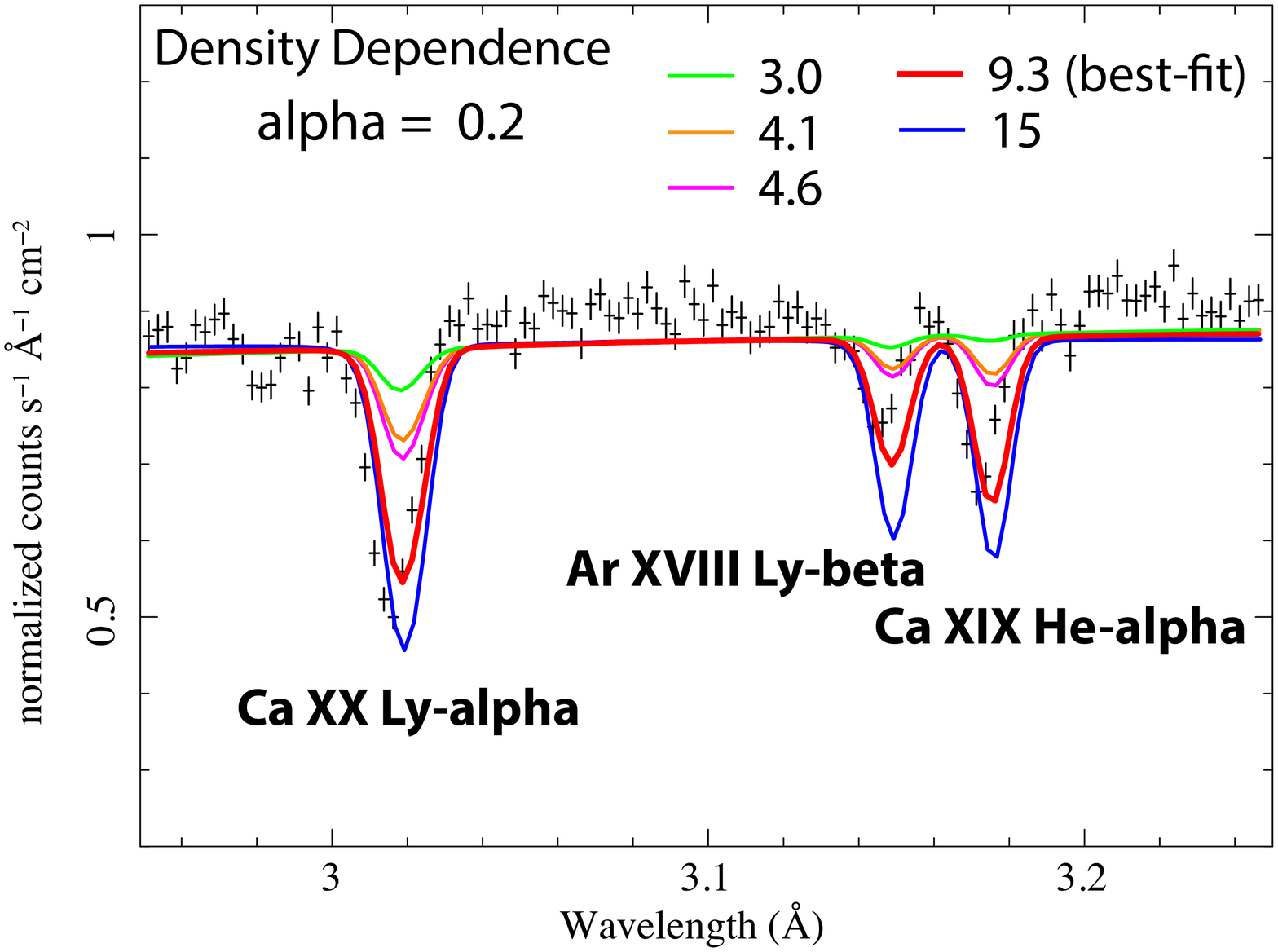}\includegraphics[trim=0in 0in 0in
0in,keepaspectratio=false,width=3.3in,angle=0,clip=false]{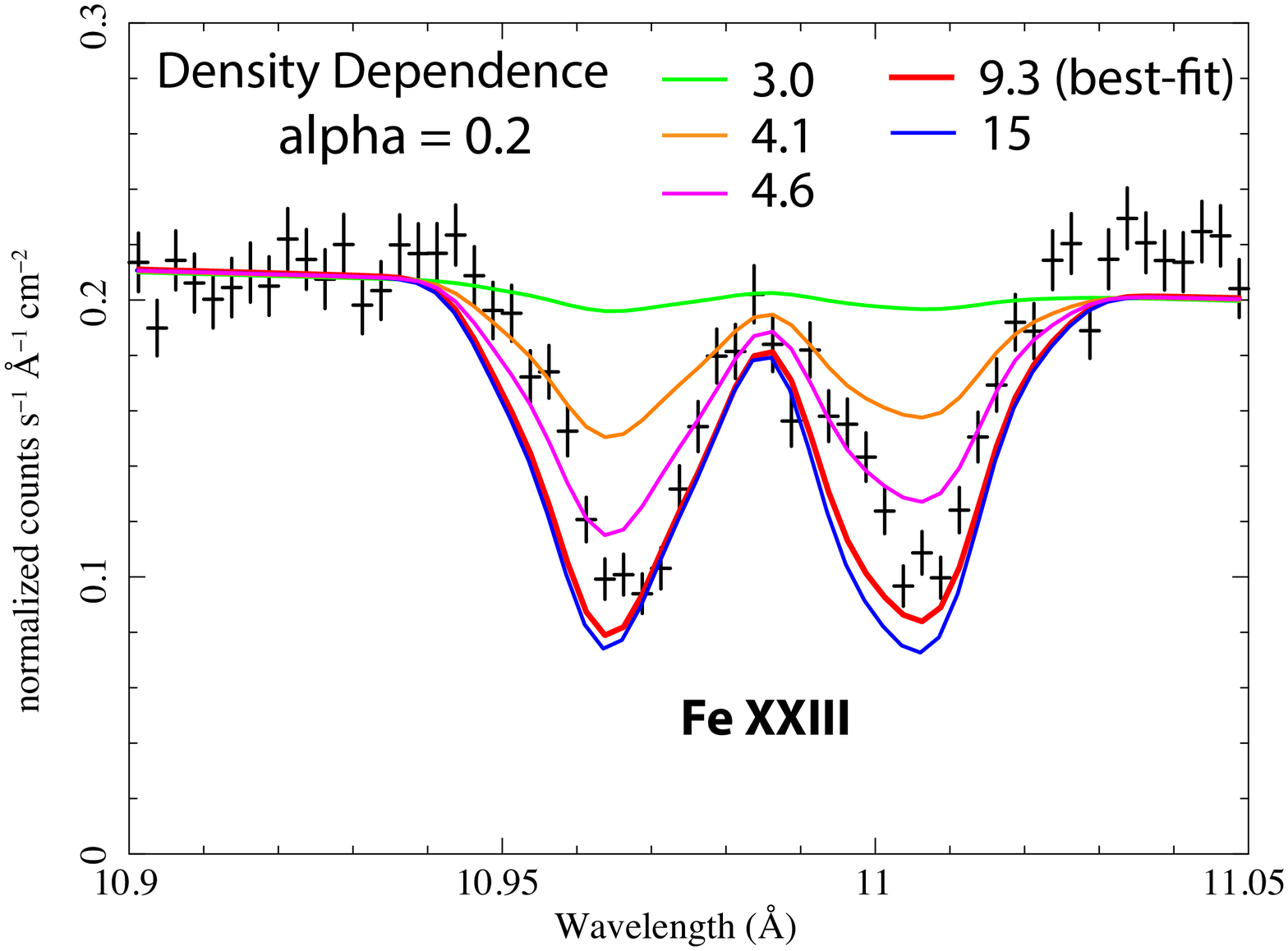}
\caption{{\bf Fig.~S2. \\ Density normalization  dependence ($\tilde{n}_{17}$) of specific lines with $\alpha=0.2$.}
Lines profiles of \caxx~Ly$\alpha$, \arxviii~Ly$\beta$, \caxix~He$\alpha$ and resonant \fexxiii, as an example, for $\alpha=0.2$ in comparison with {\it Chandra}/HETG data with $\tilde{n}_{17}=3.0$ (green), $4.1$ (orange), $4.6$ (magenta), $9.3$ (red; best-fit value) and $15$ (blue). Note that all the lines are uniquely coupled to the same density normalization $n_0 = \tilde{n}_{\rm 17} 10^{17}$ cm$^{-3}$ at the launching disk surface. It is demonstrated that the simultaneous modeling for all the observed transition lines with the MHD-wind scenario can clearly differentiate each case and favor the $\alpha=0.2$ wind of $\tilde{n}_{\rm 17}=9.3$. }
\end{center}
\end{figure}

\vskip 0pt

{\sf \noindent \textbf{Figure S2}
Density normalization  dependence ($\tilde{n}_{17}$) of specific lines with $\alpha=0.2$.

Lines profiles of \caxx~Ly$\alpha$, \arxviii~Ly$\beta$, \caxix~He$\alpha$ and resonant \fexxiii, as an example, for $\alpha=0.2$ in comparison with {\it Chandra}/HETG data with $\tilde{n}_{17}=3.0$ (green), $4.1$ (orange), $4.6$ (magenta), $9.3$ (red; best-fit value) and $15$ (blue). Note that all the lines are uniquely coupled to the same density normalization $n_0 = \tilde{n}_{\rm 17} 10^{17}$ cm$^{-3}$ at the launching disk surface. It is demonstrated that the simultaneous modeling for all the observed transition lines with the MHD-wind scenario can clearly differentiate each case and favor the $\alpha=0.2$ wind of $\tilde{n}_{\rm 17}=9.3$.}

\clearpage

\begin{figure} %=========== S.3
\begin{center}
\includegraphics[trim=0in 0in 0in
0in,keepaspectratio=false,width=6.5in,angle=0,clip=false]{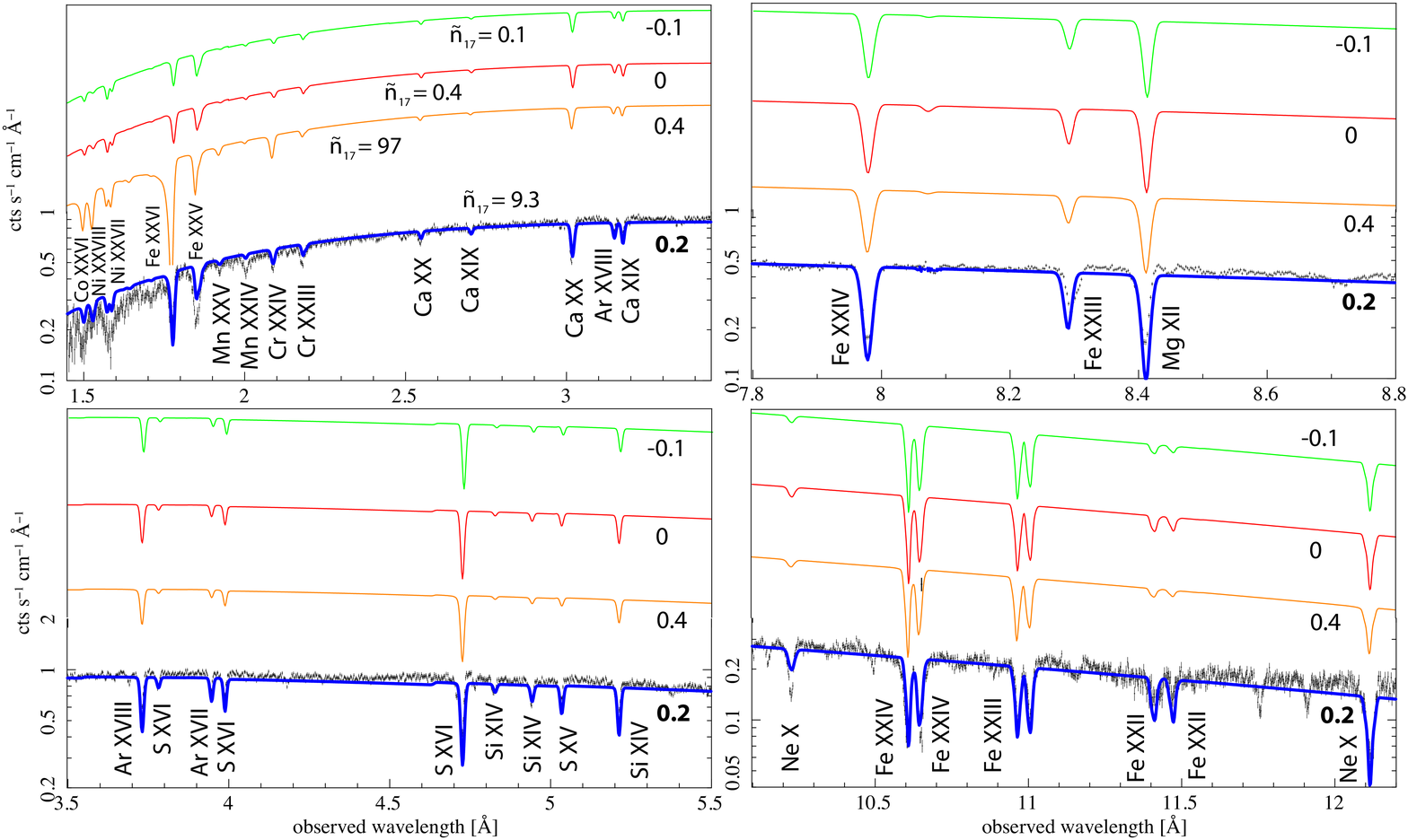}
\caption{{\bf Fig.~S3. \\ The 46-ks {\it Chandra}/HETG spectrum of \gro\  overlaid on different  global MHD-wind models of $\alpha=-0.1$ (green), $0$ (red), $0.2$ (blue; best-fit) and $0.4$ (orange).} Adopting the same baseline continuum model as in Figure~3, we examined the spectral dependence on wind density profile $\alpha$; $-0.1$ (green), $0$ (red), $0.2$ (blue; best-fit) and $0.4$ (orange) also by varying density normalization $n_o$ in each case.
Here, we find $\alpha=0.2$ (i.e. $n \propto r^{-1.2}$) to be more consistent with this broad-band data ranging from $\sim 1.5 \aa$ to $\sim 12\aa$. The overall spectral model components are the same as those in Figures~3 and 4. Note that $\alpha=-0.1, 0$ and $0.4$ model spectra  are offset by arbitrary factors in flux for presentation purpose.}
\end{center}
\end{figure}

\vskip 0pt

{\sf \noindent \textbf{Figure S3}
The 46-ks {\it Chandra}/HETG spectrum of \gro\  overlaid on different  global MHD-wind models of $\alpha=-0.1$ (green), $0$ (red), $0.2$ (blue; best-fit) and $0.4$ (orange).

\noindent Adopting the same baseline continuum model as in Figure~3, we examined the
spectral dependence on wind density profile $\alpha$; $-0.1$ (green), $0$ (red), $0.2$
(blue; best-fit) and $0.4$ (orange) also by varying density normalization $n_o$ in each
case.
Here, we find $\alpha=0.2$ (i.e. $n \propto r^{-1.2}$) to be more consistent with this broad-band data ranging from $\sim 1.5 \aa$ to $\sim 12\aa$. The overall spectral model components are the same as those in Figures~3 and 4. Note that $\alpha=-0.1, 0$ and $0.4$ model spectra  are offset by arbitrary factors in flux for presentation purpose.}

\begin{figure} %=========== fig.S4
\begin{center}
\includegraphics[trim=0in 0in 0in
0in,keepaspectratio=false,width=6.0in,angle=0,clip=false]{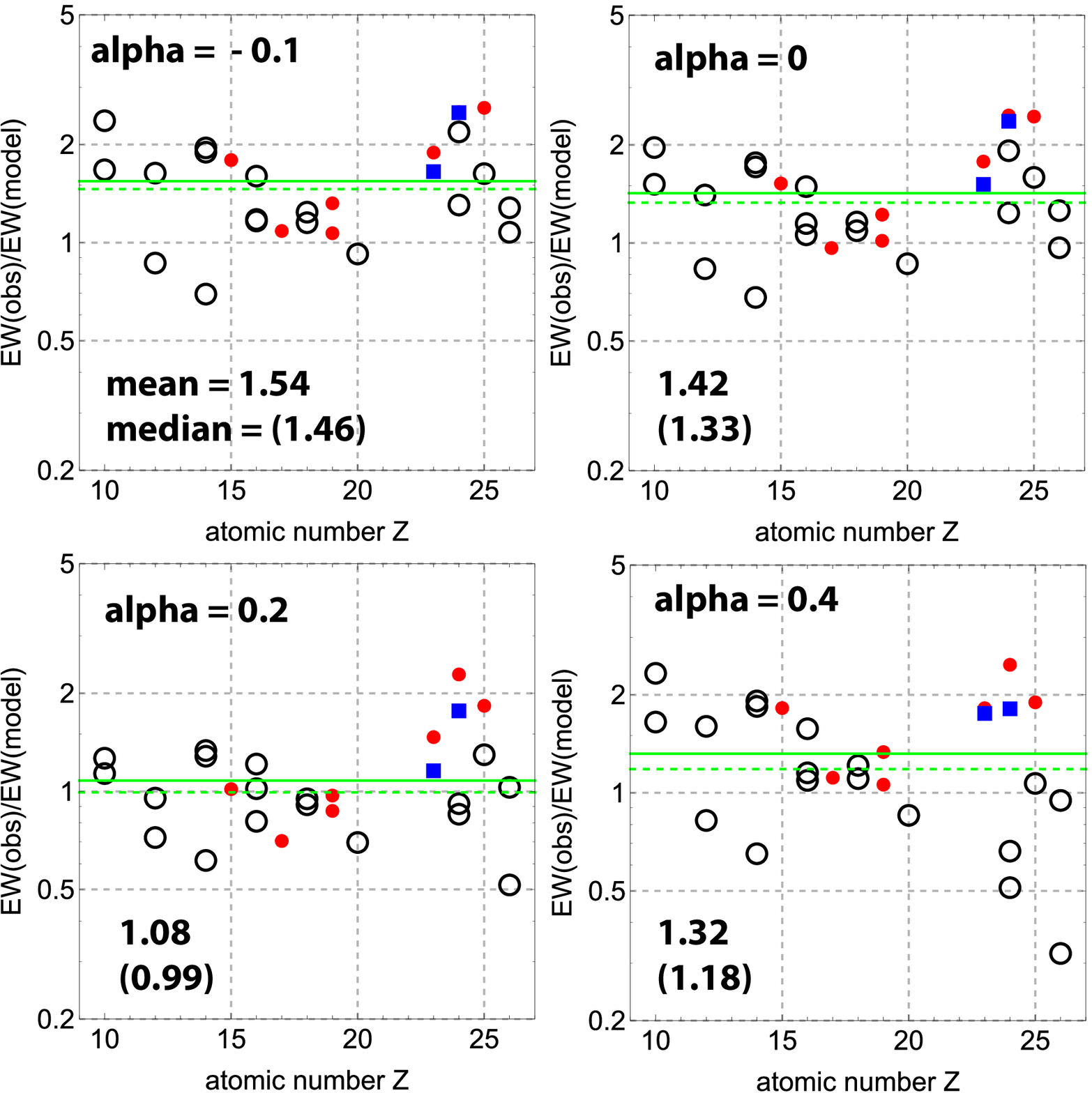}
\caption{{\bf Fig.~S4. \\ EW ratios between our model and data for $\alpha=-0.1, 0, 0.2$ and 0.4.} The ratio, EW(obs)/E(mo), is computed for major ions using models with different $\alpha$ showing its {\it mean} (solid green) and {\it median} (dotted green) values in each case (also indicated by numbers) to further support the best-fit  wind model with $\alpha=0.2$. }
\end{center}
\end{figure}

\vskip 10pt

{\sf \noindent \textbf{Figure S4}
EW ratios between our model and data for $\alpha=-0.1, 0, 0.2$ and 0.4.

\noindent The ratio, EW(obs)/E(mo), is computed for major ions using models with different $\alpha$ showing its {\it mean} (solid green) and {\it median} (dotted green) values in each case (also indicated by numbers) to further support the best-fit  wind model with $\alpha=0.2$.}

%\noindent We show the magnetic field lines (streamlines) (thick solid) along with the
%density distribution $n(r,\theta)$ (in color) and contours (thin solid) and contours of
%the ionization parameter $\log \xi$ (dashed), as a function of cylindrical distances
%$(R,Z)$ in units of the innermost launching radius $R_0$. Note the
%$\textmd{linear}(Z)-\log(R)$ character of this Figure where the line-of-sight (LoS) is
%not a straight line.}

\newpage

%-------------------- Table S1 ------------------------------
\begin{table}
\begin{flushleft}
{\bf Table S1. \\ Grids of MHD-Wind Model for \gro.} Three primary model parameters characterizing the disk-winds for spectral analysis: global density slope $\alpha$ over distance $r$ where $n(r) \propto r^{-(1+\alpha)}$, source inclination angle $\theta$ and plasma number density $\tilde{n}_{17}$ at the footpoint of the MHD wind (i.e. $n_0 \equiv \tilde{n}_{17} 10^{17}$ cm$^{-3}$).
\end{flushleft}
%\hspace{-5cm}
%\caption{{\bf Table S1. \\ Grids of MHD-Wind Model.} \\ Primary wind model parameters
%}
\vspace{0.0cm}
\centering
\begin{tabular}{l|l}
%\tablehead{Primary Parameter & Range  }
%\startdata
\hline
\\ [-0.3cm]
Primary Parameter & Range  \\ [0.1cm] \hline \hline
\\ [-0.3cm]
Density Profile $\alpha$ & -0.1, 0, 0.2, 0.4 \\ [0.2cm]
Viewing Angle $\theta$ (degrees) &  $70\deg - 80\deg$ \\ [0.2cm]
Density Normalization factor $\tilde{n}_{17}$  & 0.024 - 400 \\ [0.2cm] %heading
\hline
\hline
%\enddata
\end{tabular} \\ [0.2cm]
%$^a$ In units of $10^{17}$ cm$^{-3}$
\label{table:tab1}
\end{table}
%--------------------- end of Table 1 ------------------------

%------------------------------- Table S2
\begin{table}
\begin{flushleft}
{\bf Table S2. \\ Primary best-fit line parameters of the broad-band MHD disk-wind model for \gro.} For an $\alpha=0.2$ model with $\tilde{n}_{\rm 17}=9.3$, we present the following variables; line energy $E_o$ (in keV) and wavelength $\lambda_o$ (in $\aa$) in the source rest-frame, absorber's characteristic %distance $r_{-2}$ from the black hole in units of \sw radius of $r_s = 2.1 \times 10^{6}$ cm %where
LoS velocity $v_{\rm los}$ (in km~s$^{-1}$) at the observed trough wavelength, the range of characteristic ionization parameter $\Delta \left(\log \xi_c \right)$ corresponding to 50\% of the peak  differential column density $d N_{\rm ion}/(d \log \xi)$, the corresponding temperature range $\Delta \left(\log T_c \right)$, the characteristic line optical depth $\tau_{\rm c}$ at the peak column distance, the modeled line width EW$_{\rm mo}$ (in m$\aa$), the hydrogen-equivalent column density $N_H$ (in units of $10^{22}$ cm$^{-2}$) and the characteristic number density $n_{c}$ (in cm$^{-3}$) at the peak column distance.
\end{flushleft}
%\footnotesize
\scriptsize
%\tiny
\vspace{-0cm}
%\caption{Best-fit MHD wind characteristics for \gro. \label{tab:tab3}}
%\tablewidth{0pt}
%\tablehead{Lines & $E_{o}$$^a$ & $\lambda_{o}$$^b$  & $\log(r_{\rm -2}/r_s)$ $^c$  & $v^{\rm los}_{\rm -2}$ $^c$ &  $\log \xi_{\rm -2}$ $^c$ & $\log T_{\rm -2}$ $^c$ & $\tau_{\rm max}$ $^d$ & EW & $N_H$ $^e$ & $\log n$ $^c$ \\
%& [keV] & [$\aa$] &  &  & & &  & [m$\aa$] &  [$10^{\rm 22}$ cm$^{-2}$] &
%}
%
%\startdata
%
\centering
\begin{tabular}{lccccccccc}
\hline
\\
[-0.2cm]
Transition & $E_{o}$$^a$ & $\lambda_{o}$$^b$   & $v_{\rm los}$$^c$  &  $\Delta \left(\log \xi_c \right)$  & $\Delta \left(\log T_c \right)$  & $\tau_c$  & EW$_{\rm mo}$$^d$ & $N_H$  & $\log n_c$$^e$  \\
Line & [keV] & [$\aa$] &   [km~s$^{-1}$] & [erg~cm~s$^{-1}$] & [K] &  & [m$\aa$] &  [$10^{\rm 22}$ cm$^{-2}$] &
\\ [0.2cm] \hline \hline
%& [keV] & [$\aa$] &  &  & & &  & [m$\aa$] &  [$10^{\rm 22}$ cm$^{-2}$] &
%
\\ [-0.2cm]
\fexxvi\ Ly$\beta$ &  8.252 & 1.5028 &  320 & 3.77 - 4.91 & 6.07 - 6.86 & 0.854 & 4.27 & 28.2 & 12.2 \\
%\fexxvi\ Ly$\beta_2$ & 8.247 & 1.5034 & 3.26 & 1,553 & 5.57 & 6.96 & 0.52 & 5.92 & 28.2  & 14.2 \\
\nixxviii\ Ly$\alpha$ & 8.102 & 1.533 &  831 & 4.48 - 5.35 & 6.64 - 6.93 & 0.216 & 4.12 & 69.5 & 13.3 \\
%\nixxviii\ Ly$\alpha_2$ & 8.073 & 1.5358 & 3.09 & 1,845 & 5.69 & 6.95 & 0.257 & 7.04 & 69.5 & 14.4 \\
%\nixxvii\ He$\alpha$ & 7.805 & 1.5893 & 3.53 & 1,099 & 5.35 & 6.94 & 2.41 & 12.2 & 93.6 & 13.9 \\
%\coxxvii\ Ly$\alpha_1$ & 7.526 & 1.6473 & 3.39 & 1,305 & 5.46 & 6.95 & 0.0408 & 1.37 & 105 & 14.0 \\
%\coxxvii\ Ly$\alpha_2$ & 7.502 & 1.6527 & 3.74 & 897.7 & 5.18 & 6.92 & 0.0206 & 0.698 & 106 & 13.6 \\
%%\coxxvii\ Ly$\beta$ & \\
%\coxxvi\ He$\alpha$ & 7.242 & 1.7120 & 4.60 & 343.4 & 4.49 & 6.65 & 0.138  & 0.948 & 25.4 & 12.6  \\
%\coxxvi\ He$\beta$ & \\
\crxxiv\ Ly$\beta$ & 7.022 & 1.7662 &  2122 & 4.10 - 5.32 & 6.32 - 6.93 & 0.0381 & 11.9 & 60.3 & 12.9 \\
%\crxxiv\ Ly$\beta_2$ & 7.017 & 1.7668 & 4.52 & 484.1 & 4.55 & 6.60 & 0.0267 & 0.573 & 60.4 & 12.7 \\
\fexxvi\ Ly$\alpha$ &  6.973 & 1.7807 &  328 & 3.80 - 4.93 & 6.09 - 6.87 & 5.14 & 13.7 & 30.4 & 12.3  \\
%\fexxvi\ Ly$\alpha_2$ & 6.952 & 1.7834 & 2.58 & 3,316 & 6.11 & 6.97 & 3.27 & 19.9 & 31.2 & 15.0 \\
\fexxv\ He$\alpha$ & 6.700 & 1.8505 &  122 & 3.47 - 4.21 & 5.93 - 6.42 & 22.4 & 10.5 & 26.6 & 11.6 \\
 [0.2cm]
%\fexxiv\ Li$\alpha$ & 1.553 &  \\
%\fexxii\ & 1.5343 & 8.081 & 12.2 &   \\
%\fexxiv\ Li$\beta$ & \\
%
%
\hline
%\enddata
\end{tabular}
\vspace{0.0in}
\noindent
\label{tab:tab2}
\end{table}

%------------------------------- Table S2 continued
\begin{table}
\begin{flushleft}
{\bf Table S2. \\ Best-Fit Line Parameters of the Broad-Band MHD disk-wind model for \gro.} Continued.
\end{flushleft}
%\footnotesize
\scriptsize
%\tiny
\vspace{-0cm}
%\caption{Best-fit MHD wind characteristics for \gro. \label{tab:tab3}}
%\tablewidth{0pt}
%\tablehead{Lines & $E_{o}$$^a$ & $\lambda_{o}$$^b$  & $\log(r_{\rm -2}/r_s)$ $^c$  & $v^{\rm los}_{\rm -2}$ $^c$ &  $\log \xi_{\rm -2}$ $^c$ & $\log T_{\rm -2}$ $^c$ & $\tau_{\rm max}$ $^d$ & EW & $N_H$ $^e$ & $\log n$ $^c$ \\
%& [keV] & [$\aa$] &  &  & & &  & [m$\aa$] &  [$10^{\rm 22}$ cm$^{-2}$] &
%}
%
%\startdata
%
\centering
\begin{tabular}{lccccccccc}
\hline
\\
[-0.2cm]
Transition & $E_{o}$$^a$ & $\lambda_{o}$$^b$   & $v_{\rm los}$$^c$  &  $\Delta \left(\log \xi_{c} \right)$  & $\Delta \left(\log T_{c} \right)$  & $\tau_{\rm max}$  & EW$_{\rm mo}$$^d$ & $N_H$  & $\log n_{c}$$^e$  \\
Line & [keV] & [$\aa$] &  [km~s$^{-1}$] & [erg~cm~s$^{-1}$] & [K]  & & [m$\aa$] &  [$10^{\rm 22}$ cm$^{-2}$] &
\\ [0.2cm] \hline \hline
%& [keV] & [$\aa$] &  &  & & &  & [m$\aa$] &  [$10^{\rm 22}$ cm$^{-2}$] &
%
\\ [-0.2cm]
%\fexxvi\ Ly$\beta_1$ &  8.252 & 1.5023 & 2.99 & 2,120 & 5.77 & 6.96 & 1.04 & 19.9 & 28.2 & 14.5 \\
%\fexxvi\ Ly$\beta_2$ & 8.247 & 1.5034 & 3.26 & 1,553 & 5.57 & 6.96 & 0.52 & 5.92 & 28.2  & 14.2 \\
%\nixxviii\ Ly$\alpha_1$ & 8.102 & 1.5303 & 2.86 & 2,431 & 5.89 & 6.97 & 0.509 & 12.3 & 68.7 & 14.7 \\
%\nixxviii\ Ly$\alpha_2$ & 8.073 & 1.5358 & 3.09 & 1,845 & 5.69 & 6.95 & 0.257 & 7.04 & 69.5 & 14.4 \\
%\nixxvii\ He$\alpha$ & 7.805 & 1.5893 & 3.53 & 1,099 & 5.35 & 6.94 & 2.41 & 12.2 & 93.6 & 13.9 \\
%\coxxvii\ Ly$\alpha_1$ & 7.526 & 1.6473 & 3.39 & 1,305 & 5.46 & 6.95 & 0.0408 & 1.37 & 105 & 14.0 \\
%\coxxvii\ Ly$\alpha_2$ & 7.502 & 1.6527 & 3.74 & 897.7 & 5.18 & 6.92 & 0.0206 & 0.698 & 106 & 13.6 \\
%%\coxxvii\ Ly$\beta$ & \\
%\coxxvi\ He$\alpha$ & 7.242 & 1.7120 & 4.60 & 343.4 & 4.49 & 6.65 & 0.138  & 0.948 & 25.4 & 12.6  \\
%%\coxxvi\ He$\beta$ & \\
%\crxxiv\ Ly$\beta_1$ & 7.022 & 1.7657 & 3.79 & 834.5 & 5.14 & 6.92 & 0.0532 & 1.13 & 60.3 & 13.5 \\
%\crxxiv\ Ly$\beta_2$ & 7.017 & 1.7668 & 4.52 & 484.1 & 4.55 & 6.60 & 0.0267 & 0.573 & 60.4 & 12.7 \\
%\fexxvi\ Ly$\alpha_1$ &  6.973 & 1.7780 & 2.36 & 4,379 & 6.29 & 6.98 & 6.45 & 29.4 & 111 & 15.3  \\
%\fexxvi\ Ly$\alpha_2$ & 6.952 & 1.7834 & 2.58 & 3,316 & 6.11 & 6.97 & 3.27 & 19.9 & 31.2 & 15.0 \\
%\fexxv\ He$\alpha$ & 6.700 & 1.8505 & 3.61 & 1,026 & 5.28 & 6.93 & 34.2 & 10.8 & 26.6 & 13.8 \\
%\fexxv\ He$\beta$ & \\
%\fexxv\ Inter. & \\
\fexxiv\ Li$\alpha$ & 6.653   & 1.8635 &  1971 & 3.07 - 3.78 & 5.81 - 6.07 & 1.21 & 10.6 & 1.96 & 10.9  \\
%\fexxiv\ Li$\beta$ & 6.6167 &  \\
%\fexxiii\ Be$\alpha$ & \\
%\fexxii\ B$\alpha$ & \\
\mnxxv\ Ly$\alpha$ & 6.442 & 1.9274 &  4,068 & 4.15 - 5.43 & 6.36 - 6.94 & 0.0737 & 3.41 & 77.5 & 13.0  \\
%\mnxxv\ Ly$\alpha_2$ & 6.424 & 1.9301 & 3.67 & 956.7 & 5.24 & 6.93 & 0.0539 & 1.08 & 77.6 & 13.7 \\
%\mnxxv\ Ly$\beta_1$  & 7.625 &  \\
\mnxxiv\ He$\alpha$ & 6.186 & 2.006 &  336 & 3.31 - 4.15 & 5.87 - 6.37 & 0.168 & 3.14 & 182 & 11.4 \\
\crxxiv\ Ly$\alpha$ & 5.932 & 2.0928 &  222 & 4.12 - 5.32 & 6.34 - 6.93 & 0.238 & 5.19 & 67.1  & 12.8 \\
%\crxxiv\ Ly$\alpha_2$ & 5.917 & 2.0955 & 3.45 & 1,219 & 5.40 & 6.94 & 0.167 & 2.70 & 67.4 & 13.9 \\
\crxxiii\ He$\alpha$ & 5.68 & 2.1821 &  117 & 3.23 - 4.08 & 5.85 - 6.30 & 0.415 & 4.72 & 9.29 &  11.3 \\
\caxix\ He$\beta$ & 4.582 & 2.7054 &  182 & 2.57 - 3.42 & 5.66 - 5.91 & 0.356 & 3.42 & 10.3 & 10.4  \\
\caxx\ Ly$\alpha$ & 4.107 & 3.0185 &  25 & 3.09 - 4.09 & 5.81 - 6.30 & 0.722 & 7.60 & 14.1 & 11.2  \\
\arxviii\ Ly$\beta$ & 3.936 & 3.1507 &  190 & 2.71 - 3.92 & 5.72 - 6.17 & 0.289 & 4.44 & 5.94 & 10.8  \\
%\arxviii\ Ly$\beta_2$ & 3.934 & 3.1513 & 5.10 & 189.7 & 4.08 & 6.31 & 0.179 & 1.07 & 5.94 & 11.9 \\
\caxix\ He$\alpha$ & 3.902 & 3.1771 &  80 & 2.91 - 3.60 & 5.77 - 5.98 & 1.60 & 3.41 & 1.52 & 10.6 \\
\arxviii\ Ly$\alpha$ & 3.323 & 3.7335 &  185 & 2.84 - 3.95 & 5.76 - 6.19 & 1.62 & 11.3 & 9.87 & 10.9 \\
%\arxviii\ Ly$\alpha_2$ & 3.318 & 3.7361 & 4.56 & 353.3 & 4.52 & 6.67 & 1.12 & 3.14 & 10.0 & 12.6  \\
\sxvi\ Ly$\gamma$ & 3.276 & 3.7843 &  241 & 2.38 - 3.77 & 5.48 - 6.07 & 0.330 & 4.12 & 3.40 & 10.4 \\
\arxvii\ He$\alpha$ & 3.139 & 3.9493 &  232 & 2.73 - 3.41 & 5.73 - 5.90 & 2.02 & 5.86 & 8.23 & 10.5 \\
\sxvi\ Ly$\beta$ & 3.106 & 3.9913 &  195 & 2.39 - 3.77 & 5.49 - 6.07 & 0.950 & 7.04 & 3.48 & 10.4 \\
%\sxvi\ Ly$\beta_2$ & 3.105 & 3.9919 & 4.87 & 251.6 & 4.26 & 6.47 & 0.688 & 1.63 & 3.47 & 12.2 \\
\sxvi\ Ly$\alpha$ & 2.622 & 4.730 &  238 & 2.58 - 3.83 & 5.66 - 6.10 & 5.16 & 13.9 & 4.83 & 10.6 \\
%\sxvi\ Ly$\alpha_2$ & 2.619 & 4.7327 & 4.38 & 434.6 & 4.67 & 6.77 & 4.29 & 3.18 & 4.87 & 12.8 \\
\sixiv\ Ly$\gamma$ & 2.506 & 4.9467 &  179 & 2.09 - 3.54 & 5.19 - 5.96 & 1.38 & 5.11 & 2.15 & 9.53 \\
\sxv\ He$\alpha$ & 2.460 & 5.0387 &  295 & 2.44 - 3.24 & 5.53 - 5.85 & 8.35 & 6.58 & 0.664 & 10.1 \\ [0.2cm]
%\fexxiv\ Li$\alpha$ & 1.553 &  \\
%\fexxii\ & 1.5343 & 8.081 & 12.2 &   \\
%\fexxiv\ Li$\beta$ & \\
%
%
\hline
%\enddata
\end{tabular}
\vspace{0.0in}
\noindent
\label{tab:tab2}
\end{table}

%------------------------------- Table S2 continued
\begin{table}
\begin{flushleft}
{\bf Table S2. \\ Best-Fit Line Parameters of the Broad-Band MHD disk-wind model for \gro.} Continued.
\end{flushleft}
\scriptsize
%\footnotesize
%\tiny
%\caption{{\bf Table S2.} Continued. }
%\caption{Best-fit MHD wind characteristics for \gro. \label{tab:tab3}}
%\tablewidth{0pt}
%\tablehead{Lines & $E_{o}$$^a$ & $\lambda_{o}$$^b$  & $\log(r_{\rm -2}/r_s)$ $^c$  & $v^{\rm los}_{\rm -2}$ $^c$ &  $\log \xi_{\rm -2}$ $^c$ & $\log T_{\rm -2}$ $^c$ & $\tau_{\rm max}$ $^d$ & EW & $N_H$ $^e$ & $\log n$ $^c$ \\
%& [keV] & [$\aa$] &  &  & & &  & [m$\aa$] &  [$10^{\rm 22}$ cm$^{-2}$] &
%}
%
%\startdata
%
\centering
\begin{tabular}{lcccccccccc}
\hline
\\
[-0.2cm]
Transition & $E_{o}$$^a$ & $\lambda_{o}$$^b$  & $v_{\rm los}$$^c$ &  $\Delta \left(\log \xi_{c} \right)$  & $\Delta \left(\log T_{c} \right)$ & $\tau_{c}$ & EW$_{\rm mo}$$^d$ & $N_H$ & $\log n_c$$^e$  \\
Line & [keV] & [$\aa$] & [km~s$^{-1}$] & [erg~cm~s$^{-1}$] & [K] &  & [m$\aa$] &  [$10^{\rm 22}$ cm$^{-2}$] &
\\ [0.2cm] \hline \hline
%& [keV] & [$\aa$] &  &  & & &  & [m$\aa$] &  [$10^{\rm 22}$ cm$^{-2}$] &
%
%
\\ [-0.2cm]
\sixiv\ Ly$\beta$ & 2.376 & 5.2173 &  206 & 2.09 - 3.55 & 5.19 - 5.96 & 3.98 & 9.29 & 2.20 & 12.5 \\
%\sixiv\ Ly$\beta_2$ & 2.376 & 5.2179 &  4.87 & 251.7 & 4.26 & 6.47 & 2.17 & 1.49 & 2.20 & 12.2 \\
\sixiv\ Ly$\alpha$ & 2.006 & 6.1831 &  453 & 2.24 - 3.68 & 5.35 - 6.03 & 10.6 & 19.2 & 6.54 & 13.2 \\
%\sixiv\ Ly$\alpha_2$ & 2.000 & 6.1858 & 4.28 & 482.1 & 4.75 & 6.80 & 13.5 & 2.55 & 8.10 & 12.9 \\
%
\mgxii\ Ly$\beta$ & 1.745 & 7.1063 &  321 & 1.88 - 2.85 & 5.09 - 5.77 & 5.57 & 9.62 & 1.39 & 12.2  \\
%\mgxii\ Ly$\beta_2$ & 1.744 & 7.1069 & 5.10 & 189.7 & 4.08 & 6.31 & 3.09 & 0.732 & 1.39 & 11.9 \\
\fexxiii\ & 1.493 & 8.3001 &  319 & 2.67 - 3.27 & 5.71 - 5.86 & 7.47 & 11.7 & 1.21 & 11.7  \\
\mgxii\ Ly$\alpha$ & 1.472 & 8.4219 &  379 & 2.00 - 3.27 & 5.14 - 5.86 & 32.2 & 18.3 & 6.57 & 12.9 \\
%\mgxii\ Ly$\alpha_2$ & 1.471 & 8.4246 & 4.56 & 353.3 & 4.52 & 6.67 & 19.3 & 1.62 & 7.07 & 12.6 \\
\nex\ Ly$\beta$ & 1.211 & 10.238 &  358 & 1.63 - 2.46 & 4.86 - 5.56 & 6.64 & 8.70 & 0.901 & 11.9  \\
%\nex\ Ly$\beta_2$ & 1.210 & 10.238 & 5.29 & 154.2 & 3.93 & 6.18 & 3.54 & 0.255 & 0.900 & 11.8  \\
\nex\ Ly$\alpha$ & 1.022 & 12.134 &  513 & 1.77 - 2.54 & 5.00 - 5.64 & 36.6 & 21.2 & 1.42 & 12.6  \\
%\nex\ Ly$\alpha_2$ & 1.021 & 12.137 & 4.78 & 279.1 & 4.33 & 6.53 & 22.0 & 0.815 & 1.43 & 12.3    \\
[0.2cm]
\hline
%\enddata
\end{tabular}
\vspace{0.0in}
\noindent
%\normalsize
\begin{flushleft}
$^a$ Rest-frame energy in keV. $^b$ Rest-frame wavelength in $\aa$. $^c$ Line-of-sight (LoS) wind velocity in km~s$^{-1}$. $^d$ Model EW value. \\
$^e$ Wind number density in cm$^{-3}$.
%%$^\diamondsuit$ The value is pegged. \\
%%$^\flat$
%%$^\triangle$ The characteristic line depth where the differential ion column $dN_H/d(\log \xi)$ reaches $10^{21}$ cm$^{-2}$ along a given LoS distance.  \\
%$^c$ The characteristic value (in cgs units) where the line optical depth is $\tau=10^{-2}$.   \\
%$^d$ The maximum line optical depth. \\
%$^e$ LoS-integrated hydrogen-equivalent column density. \\
\end{flushleft}
\label{tab:tab2}
\end{table}

%\clearpage

\end{document}